\documentclass[11pt]{article}
\usepackage{jheppub}

\usepackage{bm,color,xcolor}
\usepackage{slashed} 
\usepackage{graphicx}
\usepackage{mwe}
\usepackage{adjustbox}
\usepackage[compat=1.1.0]{tikz-feynman}
\usepackage{float}
\usepackage[utf8x]{inputenc}
\usepackage[T1]{fontenc}

\usepackage{multirow}
\usepackage{tablefootnote}
\usepackage{comment}
\usepackage{mathtools}
\usepackage{appendix}
\usepackage[colorlinks=true
,urlcolor=blue
,anchorcolor=blue
,citecolor=blue 
,filecolor=blue
,linkcolor=blue
,menucolor=blue
,linktocpage=true
,pdfproducer=medialab
,pdfa=true
]{hyperref}
\newcommand{\beq}{\begin{equation}}
\newcommand{\eeq}{\end{equation}}
\newcommand{\bea}{\begin{eqnarray}}
\newcommand{\eea}{\end{eqnarray}}

\title{Heavy Neutral Leptons at Muon Colliders}

\author{Peiran Li, }
\author{Zhen Liu, }
\author{Kun-Feng Lyu}
\affiliation{\it School of Physics and Astronomy, University of Minnesota, Minneapolis, MN 55455, USA}

\emailAdd{li001800@umn.edu}
\emailAdd{zliuphys@umn.edu}
\emailAdd{lyu00145@umn.edu}

\abstract{
The future high-energy muon colliders, featuring both high energy and low background, could play a critical role in our searches for new physics. The smallness of neutrino mass is a puzzle of particle physics. Broad classes of solutions to the neutrino puzzles can be best tested by seeking the partners of SM light neutrinos, dubbed as heavy neutral leptons (HNLs), at muon colliders.
We can parametrize HNLs in terms of the mass $m_N$ and the mixing angle with $\ell$-flavor $U_\ell$. In this work, we focus on the regime $m_N > O(100)$ GeV and study the projected sensitivities on the $|U_\ell|^2 - m_N$ plane with the full-reconstructable HNL decay into a hadronic $W$ and a charged lepton. The projected reach in $|U_\ell|^2$ leads to the best sensitivities in the TeV realm.
}

\begin{document}

\maketitle
%\setcounter{tocdepth}{1}
%\tableofcontents

%\setcounter{secnumdepth}{2} 
%\setcounter{tocdepth}{2}  

%%%%%%%%%%%%%%%%%%%%%%%%%%%%%%%%%%%%%%%%%%%%%%%%%%%%%%%%%%%
%%%%%%%%%%%%%%%%%%%%%%%%%%%%%%%%%%%%%%%%%%%%%%%%%%%%%%%%%%%

\section{Introduction}
High energy muon colliders are exciting future opportunities that could probe 10~TeV scale and higher physics thoroughly thanks to its high center of mass energy and clean lepton collider environment~\cite{Boscolo:2018ytm,Delahaye:2019omf,AlAli:2021let,Black:2022cth,Narain:2022qud,Bose:2022obr,Aime:2022flm,Schulte:2022brl,Zimmermann:2022xbv}. The muon collider physics potential in the mysterious neutrino sector is yet to be understood. The Neutrino sector sources several puzzles of the Standard Model (SM), such as the origin of neutrino mass and the structure of its mixing~\cite{Bilenky:1978nj,Super-Kamiokande:1998uiq,Super-Kamiokande:1998kpq,SNO:2002tuh,KamLAND:2002uet,Bilenky:2005mx,Bilenky:2005cp,Gonzalez-Garcia:2007dlo,King:2013eh,Bilenky:2014ema,ParticleDataGroup:2022pth}. The neutrino sector can also help solve many pressing issues, such as matter-antimatter asymmetry through leptogenesis. 
Of the various approaches to account for the smallness of the neutrino mass and many puzzles, the seesaw mechanisms provide appealing natural explanations. 
In particular, heavy fermions carrying lepton numbers, singlets under the electroweak symmetry,  are introduced in a large class of seesaw models~\cite{Bjorken:1972am,Minkowski:1977sc,Sawada:1979dis,Gell-Mann:1979vob,Mohapatra:1979ia,Yanagida:1980xy,Schechter:1980gr,Chang:1985en,Langacker:1988ur,Dittmar:1989yg,Smirnov:1993af,Strumia:2006db,Mohapatra:2006gs,Kersten:2007vk,Davidson:2008bu,Kusenko:2009up,Dasgupta:2014ula} and models of (partially) composite neutrinos~\cite{Arkani-Hamed:1998wff,Grossman:2010iq,Chacko:2020zze,Cox:2021lii}. 

We dub these heavy degrees of freedom with mass above MeV as heavy neutral leptons~(HNLs)~\cite{Chanowitz:1978mv,Pilaftsis:1997jf,Ma:1998dx,Asaka:2005pn,Abdullahi:2022jlv}.
Their mass range is broad since both low-scale, and high-scale mechanisms can be viable. HNLs can mix with the ``active'' neutrinos making it potential to be detected in various experimental facilities.   

The HNL serves as a simplified benchmark elucidating a broad class of testable seesaw mechanisms. In more generalized considerations, the HNLs can uniquely serve as a singlet fermion portal between the SM sector and hidden sector physics. 
We can parametrize the HNL in terms of the mass $m_N$ and the squared mixing angle $|U_\ell|^2$ where $\ell$ refers to the flavor. Theoretically, one must introduce at least two HNLs to explain the mass difference. For simplicity of the phenomenological discussion, one chooses to turn on only one HNL at a time, which is assumed to dominate a given HNL's flavor property. One can also instead assume a fully mixed HNL, which requires proper weighting of the sensitivities from various channels. HNLs have long been the target of particle physics searches via various experimental approaches. Depending on the mass range for $m_N$, the dominant decay channel for HNL can be distinct. For example, in the mass range of GeV to W boson mass, the HNL can not decay to an on-shell massive gauge boson, so the leading decay channel is three-body final states which can enhance its lifetime~\cite{Thacker:1971hy,Helo:2010cw}. The HNL can be long-lived enough at the detector scale; hence one can exploit the displaced vertex signal to reconstruct the HNL~\cite{Helo:2013esa,Liu:2019ayx,Maiezza:2015lza,Batell:2016zod,Antusch:2016vyf,Antusch:2017hhu,Cottin:2018kmq,Abada:2018sfh,Drewes:2019fou,Bondarenko:2019tss}. While for lower mixing angles and in the sub-GeV regime, HNL can be even more long-lived and be complementarily probed at far detectors using beam-dump experiments~\cite{NA3:1985yvr,WA66:1985mfx,Gronau:1984ct,Baranov:1992vq,NOMAD:2001eyx,FMMF:1994yvb,SHiP:2015vad,Ballett:2019bgd,FASER:2018eoc,Wang:2022jff} such as CHARM~\cite{CHARM:1985nku,CHARMII:1994jjr}, NuTeV~\cite{NuTeV:1999kej}, etc.

This paper studies the HNL detection on the muon collider and focuses on the mass regime $m_N > O(100)$ GeV. The decays into heavy gauge bosons or Higgs bosons is open. Hence HNL decays promptly once produced. The study in the mass range has been conducted on the LHC~\cite{delAguila:2007qnc,Alva:2014gxa,Degrande:2016aje,Accomando:2017qcs,Pascoli:2018heg,Fernandez-Martinez:2022gsu,Abada:2022wvh,Arganda:2015ija,Ismail:2021dyp,Dev:2013wba,Das:2017nvm,Das:2017pvt,Das:2017gke,Das:2018usr,Das:2012ze} and future lepton colliders~\cite{Das:2012ze,Banerjee:2015gca,Blondel:2022qqo,Mekala:2022cmm,Chakraborty:2022pcc}. We want to stress that at the muon collider, we can achieve better constraints on the $|U_\ell|^2 - m_N$ plane, especially for the muon flavor. The first advantage of the muon collider is the clean environment compared to the hadron collider. We can easily control the background and avoid much QCD background noise. The fixed kinematics for the initial colliding muons make the events reconstruction much simpler. Furthermore, in contrast to the future proposed electron-positron colliders, muon colliders can achieve much higher center-of-mass energy (c.m.s). High Energy muon collider can run with $\sqrt{s} = 3$ TeV and 10 TeV at the first step. This can greatly improve our probe of the new physics scale or the HNL mass scale. Exploiting such significant advantages, we will show that we can open a new specific region in the parameter space and push $U_\mu^2$ down to $O(10^{-7})$ at best.

The rest of this paper is as follows. In~\autoref{sec:theory}, we make a brief introduction to the simple Type-I seesaw model and parametrize the HNL at the Lagrangian level. Then we present in detail the signal production considerations in~\autoref{sec:production}, the signal and background events generation~\autoref{sec:preselection}, and the analysis method and results in~\autoref{sec:analysis}. Finally, we conclude in~\autoref{sec:conclusion}.
%%%%%%%%%%%%%%%%%%%%%%%%%%%%%%%%%%%%%%%%%%%%%%%%%%%%%%%%%%%
%%%%%%%%%%%%%%%%%%%%%%%%%%%%%%%%%%%%%%%%%%%%%%%%%%%%%%%%%%%
%{\flushleft \bf Introduction---}
\section{Theoretical Framework}
\label{sec:theory}
%change to inverse seesaw model
This section briefly reviews the HNL theory framework relevant to this study. We begin with the simplest Type-I linear seesaw model. Introducing a new heavy fermion $N$, the Lagrangian is given by
\begin{align}
\mathcal{L}_{\nu} \supset - \lambda_\nu \bar{L} \tilde{H} N - \frac{m_N}{2} \bar{N}^c N + \text{h.c.}  \,,
\end{align}
where $\tilde{H} = i \sigma_2 H^*$. In the flavor basis $\left\lbrace \nu_L, N^c  \right\rbrace$ the mass matrix is 
\begin{align}
M_\nu = 
\left(\begin{array}{cc}
0 & m_D \\
m_D  & m_N \end{array}
\right) ,
\end{align}
where $m_D = \lambda_\nu v/\sqrt{2} $ with the vacuum expectation value of the Higgs field $v=246~$GeV. Diagonalizing the mass matrix, we get the two eigenvalues of the masses
\begin{align} 
m_\nu \equiv m_1  \simeq  \frac{m_D^2}{m_N}  =  m_D \sin \theta, \quad
m_2  \simeq m_N + \frac{m_D^2}{m_N} \simeq m_N
\end{align}
where we have assumed $m_N \gg m_D$. The mass eigenstates are the mixing of the two flavor states with mixing angle $\sin \theta = m_D/ m_N$ 
\begin{equation}
\nu_m = \cos\theta \nu_L + \sin\theta N_R^c \ , \quad\quad
N_m = \cos\theta N_R -  \sin\theta  \nu_L^c   \ .
\end{equation}
Here the subscript $m$ refers to the mass eigenstate. In terms of physical mass, we can express
\begin{equation}
|U_\ell|^2 = \sin^2 \theta = \dfrac{m_\nu}{m_N}  \ ,
\end{equation}
where we have converted to the common convention $U_\ell$ to describe the mixing angle. For $\lambda_\nu$ taking $O(1)$ value, the mass of the HNL is too large to be accessible at colliders. There are more extended seesaw models incorporating lighter HNLs which can be testable. One example is the inverse seesaw model~\cite{Das:2012ze,Dias:2012xp,Law:2013gma,Mohapatra:1986bd,Ma:1987zm,Ma:2009gu,Bazzocchi:2010dt}. Both left-handed and right-handed neutrinos are introduced as $F_L$ and $F_R$, respectively. The relevant Lagrangian terms are
\begin{equation}
-\mathcal{L}_\nu \supset   \lambda_\nu \bar{L} \tilde{H} F_R + M \bar{F}_L F_R  + \dfrac{1}{2} \mu \bar{F_L^c} F_L + \text{h.c.}
\label{eq:inverseLag}
\end{equation}
The first two terms are the Dirac mass terms, and the third is the Majorana mass term. In the basis of $(\nu_L, \,\, F_R^c, \,\, F_L)$, the mass matrix is given by
\begin{equation}
M_\nu = \begin{pmatrix}
0 &  m_D  &  0 \\
m_D  &  0  &  M \\
0  &  M  &  \mu
\end{pmatrix}
\end{equation}
where again we have $m_D = \lambda_\nu v/\sqrt{2} $.  Under the limit $\mu \ll m_D \ll M$, the three eigenvalues of masses are given by
\begin{equation}
m_\nu \approx \dfrac{m_D^2}{M} \dfrac{\mu}{M} \,  \quad
M_{1,2} \approx M \pm \dfrac{1}{2} \mu
\end{equation}
and the mixing angle is given by
\begin{equation}
|U_\ell|^2 = \sin^2\theta = \left( \dfrac{m_D}{M} \right)^2 =\dfrac{m_\nu}{\mu}  \ .
\end{equation}
Since $\mu$ is a free parameter that is far smaller than the HNL mass, the mixing angle may be sizeable. Another interesting seesaw model is the linear seesaw model~\cite{Wyler:1982dd,Akhmedov:1995ip,Akhmedov:1995vm}, where we introduce extra fermion with Dirac mass $m_\psi$, and the mixing angle is given by
\begin{equation}
\sin\theta = \dfrac{m_\nu}{m_\psi}   \ ,
\end{equation}
The presence of $m_\psi$ violates the lepton number. Hence, we can treat $m_N$, the mass of HNL, and the squared mixing angle, $|U_\ell|^2$, as free parameters. As discussed in the introduction section, we choose to turn on one flavor each time and introduce one HNL which can be either Dirac or Majorana.
Such a choice is convenient for collider phenomenology analysis, and we can focus on a specific flavor. Under the mixing angle with the SM neutrino flavor, we can write down the relevant interactions
\begin{equation}
\begin{split}
\mathcal{L} \supset 
&\dfrac{g U_\ell}{\sqrt{2}} \left(W_\mu \bar{l}_L \gamma^\mu N + \text{h.c.} \right)  -\dfrac{g U_\ell}{2 \cos\theta_w} Z_\mu\left(\bar{\nu}_L \gamma^\mu N + \bar{N} \gamma^\mu \bar{\nu}_L \right) 
 - U_\ell \dfrac{m_N}{v} h \left(\bar{\nu}_L  N + \bar{N} \nu_L \right)  \,
\end{split}
\end{equation}
where we only keep the linear term of $U_\ell$. The HNL couples to massive gauge bosons and Higgs bosons with the size suppressed by the mixing angle $U_\ell$. For $m_N$ above $O({\rm GeV})$ scale, the main production channel for HNL is from the on/off-shell massive gauge boson decay. The produced HNLs are fully polarized due to the mixing with SM left-handed neutrinos. We focus on the mass range $m_N$ larger than $O(100)$ GeV, the dominant decay channel for HNL is to on-shell massive gauge boson and Higgs boson. These are 1 to 2 processes so HNL will promptly decay after it is produced at the interaction point. For $m_N \gg m_h$, the lifetime of HNL is given by~\cite{Pascoli:2018heg} 
\begin{equation}
    \Gamma_N \approx \dfrac{g^2 m_N^3}{32 \pi^2 m_W^2} |U_\ell|^2
\end{equation}
%We show its lifetime in \autoref{fig:HNL_lifetime}. 

Whether the HNL is Dirac or Majorana does not affect the production signature,  only the left-handed component of HNL shows up due to mixing. However, it indeed impacts the decay pattern. As is shown in~\cite{deGouvea:2021ual,Balantekin:2018ukw,BahaBalantekin:2018ppj}, for the decay channel $N \rightarrow \nu +X$ in which X is a self-conjugate boson, one can define the following forward-backward symmetry $A_{\rm FB}$ in the rest frame of $N$
\begin{equation}
A_{\rm FB} \equiv \dfrac{ \int_0^1 \frac{d \Gamma}{d \cos \theta_X} d \cos \theta_X -  \int_{-1}^0 \frac{d \Gamma}{d \cos \theta_X} d \cos \theta_X }{\int_0^1 \frac{d \Gamma}{d \cos \theta_X} d \cos \theta_X +  \int_{-1}^0 \frac{d \Gamma}{d \cos \theta_X} d \cos \theta_X}
\end{equation}
For the Majorana HNL, the decay is isotropic at the leading order, and hence $A_{\rm FB}$ is vanishing. While for Dirac spinor, $A_{\rm FB}$ is non-zero due to the helicity selection. A similar case is for the charged lepton final states.
For the Dirac spinor case, $N_\ell$ can only decay to $\ell^-$ with the same lepton number plus $W^+$. While for the Majorana spinor case, where the lepton numbers are no longer conserved, both $(\ell^-, W^+)$ and $(\ell^+, W^-)$ can appear in the final states. 
One can show that the sum of angular distributions will be isotropic if $N$ is Majorana fermion and non-isotopic for the Dirac HNL case. In this study, we only show the results of kinematics simulating the Dirac fermion case, and one can further optimize for the Majorana case.

%%%%%%%%%%%%%%%%%%%%%%%%%%%%%%%%%%%%%%%%%%%%%%%%%%%%%%%%%%%
%%%%%%%%%%%%%%%%%%%%%%%%%%%%%%%%%%%%%%%%%%%%%%%%%%%%%%%%%%%
\section{HNL Productions at High Energy Muon Colliders}
\label{sec:production}

We study two benchmark future muon collider running scenarios. The 3 TeV muon collider benchmark has a 3~TeV center of mass energy and $1~\text{ab}^{-1}$ integrated luminosity. The 10 TeV muon collider benchmark has a 10~TeV center of mass energy with $10~\text{ab}^{-1}$ integrated luminosity.

We divide our simulation and analysis into the following four categories: the $\mu$-flavored and $e$-flavored HNL at $\sqrt{s} = 3$ TeV and 10 TeV. Note that the $\tau$-flavored HNL results will be similar to that of the $e$-flavored HNL. We will discuss an extrapolation in the final section when we show the results. The Muon collider is advantageous for the muon-flavor HNL production, as the dominant production channel for $N_\mu$ is $t$-channel which avoids the $1/s$ suppression for the $s$-channel processes. While for $e$-flavored case, $N_{e}$ are mainly produced via the $s$-channel at 3 TeV and Vector Boson Fusion (VBF) processes at 10 TeV. 

\begin{figure}[th]
    \centering
    \includegraphics[width=5cm]{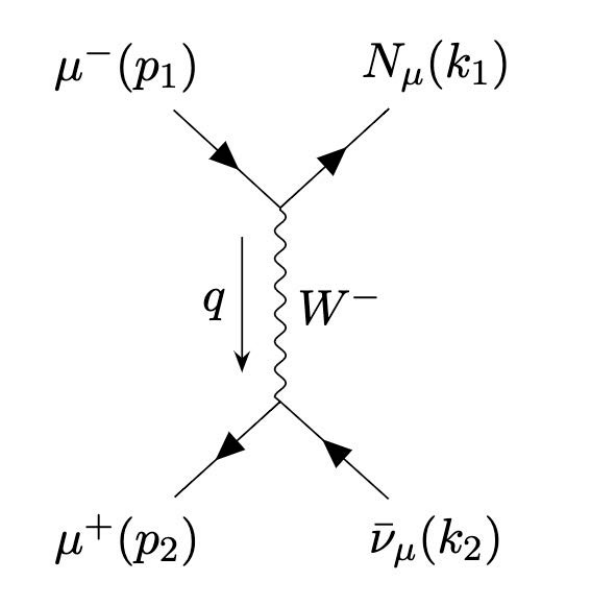}
    \qquad
    \includegraphics[width=6cm]{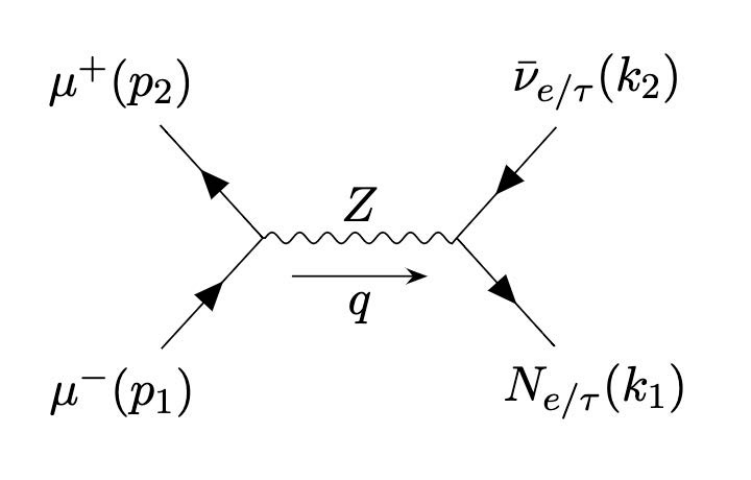}
    \caption{The Feynman diagrams for $\mu^+\mu^- \rightarrow N_\ell + \bar{\nu}_\ell$. The left panel is for $N_\mu$ and the right panel is for $N_{e/\tau}$.}
    \label{fig:feynman_diagram}
\end{figure}

For the muon-flavored HNL, the Feynman diagram for the leading $t$-channel process is shown in the left panel of  \autoref{fig:feynman_diagram}. By exchanging $t$-channel $W$ boson, initial state muon pairs produce pairs of neutrinos, one of which can be $N_\mu$ due to mixing. At high $\sqrt{s}$, the forward scattering dominates; hence $N_\mu$ is emitted almost along the direction of $\mu^-$ while $\bar{N}_{\mu}$ comes out backward. The forward scattering dominance becomes more and more prominent for lighter and lighter HNLs.

\begin{figure}[th]
    \centering
    %\subfloat[\centering e/tau]
    \includegraphics[width=7cm]{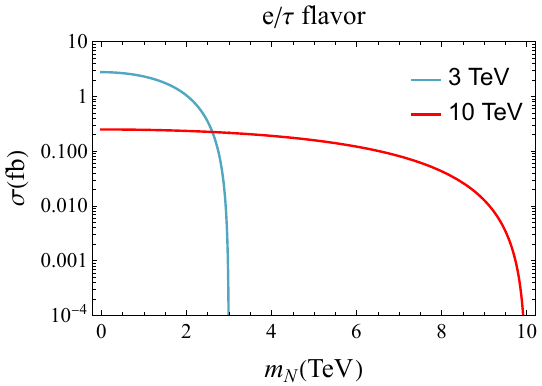}
    \qquad
    %\subfloat[\centering mu]
    \includegraphics[width=6.65cm]{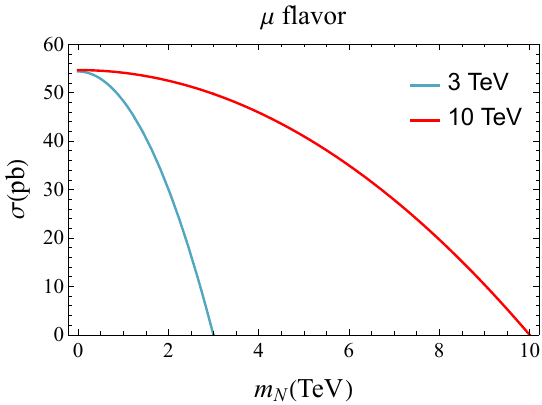}
    \caption{The cross section of the 2-to-2 process $\mu^+ \mu^- \rightarrow N_\ell + \bar{\nu}_\ell$ as a function of $m_N$ for $\ell = \mu$ and $e$ respectively. The blue curve refers to $\sqrt{s} = 3$ TeV and the red curve is for $\sqrt{s} = 10$ TeV.}
    \label{fig:2-2_xs}%
\end{figure}

While for $e/\tau$-flavored production, the leading $s$-channel diagrams are shown in the right panel of \autoref{fig:feynman_diagram}. The muon pair first annihilates into an off-shell Z boson, which converts into $\nu_{e} + \bar{N}_{e}$ or $\bar{\nu}_e + N_e$, and this process is $p$-wave angular distributed. The angular distribution is more evenly distributed in contrast to the forward production dominance from the muon flavor. However, its cross-section is highly suppressed by the $s$-channel propagator. We plot the leading 2-to-2 cross section in \autoref{fig:2-2_xs} for comparison.

Another important contribution to HNL production is the VBF processes. As elaborated in~\cite{Han:2020uid,AlAli:2021let}, with the energy scale increased, for a fixed HNL mass, the role of the virtual electroweak gauge bosons becomes more and more relevant. The growth of the VBF rate requires a resummation of large logarithms appearing in the process. By perturbatively computing the splitting function, we can use the Parton Distribution Function (PDF)~\cite{Kane:1984bb,Dawson:1984gx,Han:2020uid,Fornal:2018znf,Chen:2016wkt,AlAli:2021let,Costantini:2020stv} to compute the scattering of these off-shell gauge bosons. Naively the cross section scales like $\alpha_W^2 \log(s)$ where $\alpha_W = g_W^2/(4\pi)$ refers to the electroweak coupling square. Its ratio to the $s$-channel cross-section can be estimated by~\cite{AlAli:2021let}
\begin{equation}
    \dfrac{\sigma_{\rm VBF}}{\sigma_{\rm s-channel}} \propto
    \alpha_W^2 \dfrac{s}{m_Z^2} \log^2 \dfrac{s}{m_W^2} \log \dfrac{s}{m_N^2}
\end{equation}

As the muon collider center of mass energy increases, the contribution from the VBF process can dominate over the $s$-channel process. The next section will show that, at $\sqrt{s} = 3$ TeV, the cross-section of the VBF process is sub-dominant with a comparable size. While $\sqrt{s}$ reaches up to 10 TeV, the VBF process primarily contributes to the total signal rate for $e$-flavored HNL. 

\begin{figure}[th]
    \centering
    %\subfloat[\centering e/tau]
    \includegraphics[width=9cm]{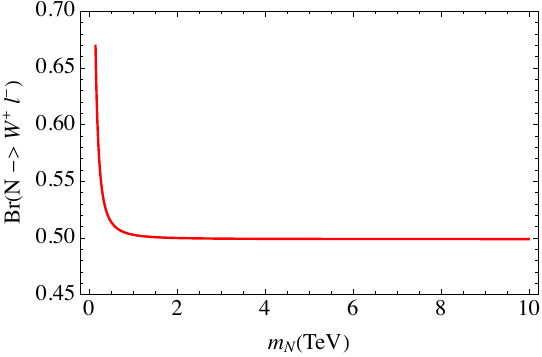} %
    \caption{The branching ratio of the Dirac HNL decay channel $N_\ell \longrightarrow W^+ \ell^-$ as a function of $m_N$. This channel allows full reconstruction of the HNL decay and reaches the equivalence limit of 50\% at $\sim1$~TeV.}
    \label{fig:HNL-decay-br}%
\end{figure}

We used both analytical calculation, convolution with EW PDF, and event generator to study and cross-check the signal rates. We simulate the whole process using {\tt MadGraph5\textunderscore aMC$@$NLO}~\cite{Alwall:2014hca,Ruiz:2021tdt} to generate the signal events. The Universal FeynRules Output (UFO)~\cite{Degrande:2011ua} models {\tt HeavyN}~\cite{Alva:2014gxa,Degrande:2016aje,Atre:2009rg} generated by FeynRules~\cite{Degrande:2011ua,Alloul:2013bka} is employed. The charged current decay of  HNL is the focus of this study due to enhanced observability through the charged leptons and the possibility of reconstructing HNL resonance from the hadronic $W$ decays. 
The branching fraction of HNL decaying into this channel is shown in \autoref{fig:HNL-decay-br} as a function of $m_N$. The branching fraction is independent of $|U_\ell|^2$ to the leading order. 
The backgrounds will be dominantly from various SM processes with hadronic $W$ bosons and as well hadronic $Z$ bosons. 
The next section presents the pre-selection of the signal and background events for the four categories.

%%%%%%%%%%%%%%%%%%%%%%%%%%%%%%%%%%%%%%%%%%%%%%%%%%%%%%%%%%%
%%%%%%%%%%%%%%%%%%%%%%%%%%%%%%%%%%%%%%%%%%%%%%%%%%%%%%%%%%%
\section{Signal and Background after Pre-Selection Cuts}
\label{sec:preselection}

This section discusses the baseline signal and background cross-sections after the pre-selection cuts (PSC). As we shall see later, pre-selection cuts are sometimes needed for a consistent definition of the cross-section. PSC can help identify backgrounds that can fake the signal with particles along the beampipe or the shielding region. Also, it can help physically regular a few singularities we could encounter. We will elaborate in a later part of this section. We list all possible production channels for the signal, including the leading 2-to-2 and 2-to-4 (VBF) channels. As mentioned above, the final states considered are hadronic weak-boson jets plus a charged lepton. We assume the $W$ and $Z$ boson can be constructed by the dijets or merged fat jet system, with a conservative assumption that we cannot distinguish between hadronic $W$ and hadronic $Z$. Consequently, for signal we first consider the chanel $\mu^+\mu^- \rightarrow \bar{\nu}_\ell + N_\ell$, followed by $N_\ell \rightarrow \ell^- + W^+$. For background, we generate $\mu^+\mu^- \rightarrow V + \ell^- + X$ in which $V$ refers to either $W$ or $Z$ gauge boson, and $X$ refers to other particles escaped detection. In what follows, we discuss them channel by channel.

Several subtle points in the computation of the VBF processes are worth mentioning. First, the emitted virtual photon from the incoming muon has infrared divergence. Such divergence can be avoided using the equivalent photon approximation (EPA) or the photon PDF. The simple EPA needs to be replaced by consistent electroweak (EW) treatment at high energy muon colliders~\cite{Han:2020uid}. In principle, inclusive VBF processes are more conveniently calculated using PDFs of EW gauge bosons.\footnote{A matching procedure between the massless splitting functions and massive ``partons'' needs to be cautiously made.}
However, when we compute the partonic cross section of the gauge bosons scattering, $t$-channel singularities arise as PDF assumes the ``incoming'' gauge bosons are on-shell, while in reality, they should be having space-like-separations of off-shell momentum. In other words, we encounter a fake singular behavior caused by the quasi-real partonic assumption. We will analyze the individual channel carefully and extract the primary contribution to avoid such an unphysical issue.
Another interesting issue that we plan to elaborate more in future work arises. 
Some of the processes can be seen as the scattering of the decay debris of muon with the other anti-muon. So in the 2-to-4 diagrams, we encounter the on-shell internal propagator, which is the so-called $t$-channel singularity~\cite{Ginzburg:1995bc,Dams:2002uy,Melnikov:1996iu}. It is claimed that the cross section of such diagrams is proportional to the incoming muon beam size, so it is not a concern for us. We will elaborate in the following when we deal with the explicit diagrams.

Since the final states are composed of charged lepton, neutral lepton, and gauge bosons, we define the pre-selection cuts (PSC) on the charged lepton to be visible as follows:
\begin{equation}
    p_T(\ell)>20~\text{GeV}, \quad
    |\eta(\ell)|<2.5
\end{equation}

In the following subsections, we list the cross section of various channel processes. It should be emphasized that we only enumerate the $N_\ell$ production, and the $\bar{N}_\ell$ production is the charge conjugate process of that for $N_\ell$. The same cases apply to the background events. We double its rate and the values in the rate columns have included both the positive charge and minus charge cases.

%%%%%%%%%%%%%%%%%%%%%%%%%%%%%%%%%%%%%%%%%%%%%%%%%%%%%%%%%%%
\subsection{$\mu$-flavored HNL at 3 TeV}
We list the signal cross section for various channels in \autoref{Table:mu_flavor_3TeV_signal}. The mass $m_N$ is set as 1~TeV as a benchmark value. The value of the cross-section $\sigma/|U_\ell|^2$ is after $N_\ell$ decaying into $W^++\ell^-$ and a pre-selection cut on $\ell^-$ discussed earlier. The first row shows the result of the leading 2-to-2 $t$-channel process, which is over 40 pb. The bottom rows are for the VBF channels where the muon pair in the final states are made invisible. As we can see, the signal rate of the leading $t$-channel process is at least two orders larger in magnitude than that of the 2-to-4 VBF processes. Hence we only keep the 2-body process to when considering the muon-flavored HNL.

\begin{table}[H]
\centering
\resizebox{\textwidth}{!}{%
\begin{tabular}{|c|c|c|c|c|}
\hline
Type & Signal process & \multicolumn{1}{c|}{\begin{tabular}[c]{@{}c@{}}$\sigma/|U_\mu|^2$ (w. conj. channel\tablefootnote[2]{When we report cross-sections for the rest of this paper, we include both processes and their charge conjugate processes. The reason to label the charges clearly is to avoid ambiguities of various cuts, particularly regarding collinear particles.}) 
 \\$m_N=1~\text{TeV}$\end{tabular}}  & Pre-selection cut (PSC) & Included\\
\hline
$t$-channel & $\mu^+ \mu^- \longrightarrow N_\mu \Bar{\nu}_\mu$ & 43.5 pb & PSC & Yes\\
VBF & $\mu^+ \mu^- \longrightarrow \mu^+ \mu^- N_\mu \Bar{\nu}_\mu$ & $\sim 1$ pb & -- & No\\
VBF & $\mu^+ \mu^- \longrightarrow \Bar{\nu}_\mu \nu_\mu N_\mu \Bar{\nu}_\mu$ & $\sim 0.1$ pb & -- & No\\
\hline
\end{tabular}%
}
\caption{The signal rate for $N_\mu$ at 3 TeV muon collider. The pre-selection cut is defined as: $p_T(\ell^-)>20~\text{GeV}$ and $|\eta(\ell^-)|<2.5$ where $\ell^-$ comes from $N_\ell$ decay. The cross-section includes the charge conjugate process.
}
\label{Table:mu_flavor_3TeV_signal}
\end{table}

Various channels for the background events are listed in \autoref{Table:mu_flavor_3TeV_bkg}. The primary channels are $\mu^+ \mu^- \longrightarrow W^+ \mu^- \Bar{\nu}_\mu$ with $\sigma \sim 0.8$ pb and $\mu^+ \mu^- \longrightarrow Z \mu^+ \mu^-$ with $\sigma \sim 0.6$ pb. The sub-leading VBF processes are listed in the bottom lines, and we will analyze them in detail next.

The cross-sections of each background process are shown below. The pre-selection cut is defined as: $p_T(\mu^-)>20~\text{GeV},|\eta(\mu^-)|<2.5$

\begin{table}[H]
\centering
\resizebox{\textwidth}{!}{%
\begin{tabular}{|c|c|c|c|c|}
\hline
Type & Background process & $\sigma$ (w. conj. channel)& Pre-selection cut (PSC) & Included\\
\hline
$t$-channel & $\mu^+ \mu^- \longrightarrow W^+ \mu^- \Bar{\nu}_\mu$ & 0.788 pb & PSC & Yes\\
\hline
$t$-channel & $\mu^+ \mu^- \longrightarrow Z \mu^+ \mu^-$ & $0.62$ pb & PSC \& missing $\mu^+$ & Yes\\
\hline
VBF & $\mu^+ \mu^- \longrightarrow \mu^+ \mu^- W^+ \mu^- \Bar{\nu}_\mu$ & $0.18$ pb & PSC \& missing $\mu^+ \mu^-$ & Yes\\
\hline
VBF & $\mu^+ \mu^- \longrightarrow \Bar{\nu}_\mu \nu_\mu W^+ \mu^- \Bar{\nu}_\mu$ & $0.028$ pb & PSC & Yes\\
\hline
VBF & $\mu^+ \mu^- \longrightarrow \Bar{\nu}_e \nu_e W^+ \mu^- \Bar{\nu}_\mu$ & $0.004$ pb & PSC & No\\
\hline 
\end{tabular}%
}
\caption{$N_\mu$ background at 3 TeV. The cross section includes the charge conjugate process.}
\label{Table:mu_flavor_3TeV_bkg}
\end{table}

For the first VBF process $\mu^+ \mu^- \longrightarrow \mu^+ \mu^- W^+ \mu^- \Bar{\nu}_\mu$, the major contribution is from the virtual photon emitted from the muon, namely from the following two partonic subprocesses: 
\begin{align*}
    \gamma^* \gamma^* & \longrightarrow   W^+ W^-,\quad W^- \longrightarrow  \mu^- \Bar{\nu}_\mu\\
    Z^* \gamma^* & \longrightarrow  W^+ W^-,\quad W^- \longrightarrow  \mu^- \Bar{\nu}_\mu   \  .
\end{align*}
We extract the leading contribution, namely the pair of $\mu^-\bar{\nu}_\mu$ from the on-shell decay of $W^-$ boson. Other diagrams with off-shell propagators will be suppressed. For the $Z^*$ initial state, if we choose to generate $Z \gamma \rightarrow W^+ \mu^- \bar{\nu}_\mu$, we will encounter diagrams with the on-shell internal muon propagator,  the unphysical $t$-channel singularity. As we can see in the 2-to-5 process reported, this diagram is suppressed and can be ignored. We can focus on the two produced on-shell $W$ bosons followed by the decay of $W^-$. 

For the second VBF process, $\mu^+ \mu^- \longrightarrow \bar{\nu}_\mu \nu_\mu W^+ \mu^- \Bar{\nu}_\mu$, it is dominated by the PDF of W boson. Here we need to make sure $\mu^-$ is visible, which ensures that the photon PDF does not contribute significantly and there is no singularity for this channel. To avoid fake singularity when using $W$ boson PDF, we choose to run the full 2-to-5 process. Its cross-section is around 15\% of the first VBF processes with exchanging photons. This is consistent with our physics intuition since the photon PDF dominates over the massive gauge bosons. 
The last VBF process is far more suppressed because $\bar{\nu}_e \nu_e$ pair is mainly from an on-shell $Z$ boson, and the VBF processes are suppressed due to the pre-selection cuts on $\mu^-$. We elect to ignore this channel due to its tiny contribution.

%%%%%%%%%%%%%%%%%%%%%%%%%%%%%%%%%%%%%%%%%%%%%%%%%%%%%%%%%%%
\subsection{$\mu$-flavored HNL at 10 TeV}

Here we list the cross section in \autoref{Table:mu_flavor_10TeV_signal} and \autoref{Table:mu_flavor_10TeV_bkg} for signal and background channels on 10 TeV muon collider. The strategy is similar to the 3~TeV case discussed earlier.

\begin{table}[H]
\centering
\resizebox{\textwidth}{!}{%
\begin{tabular}{|c|c|c|c|c|}
\hline
Type & Signal process & \multicolumn{1}{c|}{\begin{tabular}[c]{@{}c@{}}$\sigma/|U_\mu|^2$ (w. conj. channel)\\$m_N=1~\text{TeV}$\end{tabular}} & Pre-selection cut (PSC) & Included\\
\hline $t$-channel & $\mu^+ \mu^- \longrightarrow N_\mu \Bar{\nu}_\mu$ & 20.28 pb & PSC & Yes\\
VBF & $\mu^+ \mu^- \longrightarrow \mu^+ \mu^- N_\mu \Bar{\nu}_\mu$ & $\sim 1$ pb & -- & No\\
VBF & $\mu^+ \mu^- \longrightarrow \Bar{\nu}_\mu \nu_\mu N_\mu \Bar{\nu}_\mu$ & $\sim0.1$ pb & -- & No\\
\hline 
\end{tabular}%
}
\caption{The signal rate for $N_\mu$ at 10 TeV. The cross section includes the charge conjugate process.}
\label{Table:mu_flavor_10TeV_signal}
\end{table}

We ignore the VBF processes for signal only to keep the leading $t$-channel diagram. The signal rate before pre-selection cuts increases slightly compared to that on 3 TeV.

For background events, the cross-section of the $t$-channel processes is reduced compared to 3 TeV. This is because most of the forwarding events are rejected by the PSC. The cross-section of VBF processes is increased by at least a factor of 2. The gauge boson PDF will be enhanced as $\sqrt{s}$ jumps from 3 TeV to 10 TeV. On the other hand, the final state particles are more boosted, leading to fewer events passing the pre-selection cuts. As a result, the VBF process is comparable in size to that of 2-to-3 $t$-channel processes.  

\begin{table}[H]
\centering
\resizebox{\textwidth}{!}{%
\begin{tabular}{|c|c|c|c|c|}
\hline
Type & Background process & $\sigma$ (w. conj. channel)& Pre-selection cut (PSC) & Included\\
\hline
$t$-channel & $\mu^+ \mu^- \longrightarrow W^+ \mu^- \Bar{\nu}_\mu$ & 0.214 pb & PSC & Yes\\
\hline
$t$-channel & $\mu^+ \mu^- \longrightarrow Z \mu^+ \mu^-$ & $0.464$ pb & PSC \& missing $\mu^+$ & Yes\\
\hline
VBF & $\mu^+ \mu^- \longrightarrow \mu^+ \mu^- W^+ \mu^- \Bar{\nu}_\mu$ & $0.401$ pb & PSC \& missing $\mu^+ \mu^-$ & Yes\\
\hline
VBF & $\mu^+ \mu^- \longrightarrow \Bar{\nu}_\mu \nu_\mu W^+ \mu^- \Bar{\nu}_\mu$ & $0.0686$ pb & PSC & No\\
\hline 
\end{tabular}%
}
\caption{$N_\mu$ background at 10 TeV. The cross section includes the charge conjugate process.}
\label{Table:mu_flavor_10TeV_bkg}
\end{table}

%%%%%%%%%%%%%%%%%%%%%%%%%%%%%%%%%%%%%%%%%%%%%%%%%%%%%%%%%%%
\subsection{$e$-flavored HNL at 3 TeV}

Compared to the $\mu$-flavored case, there is no $t$-channel enhancement for the $e$-flavored case. The leading 2-to-2 channel is via exchanging $Z$ boson in $s$-channel. Hence its signal rate scales like $s^{-1}$. The leading VBF processes are similar to the $\mu$-flavored case, except that the $N_e$ cannot be emitted directly from the muon.

\begin{table}[H]
\centering
\resizebox{\textwidth}{!}{%
\begin{tabular}{|c|c|c|c|c|c|}
\hline
\multirow{2}{*}{Type} & \multirow{2}{*}{Signal process} & \multicolumn{2}{c|}{$\sigma/|U_e|^2$ (w. conj. channel)} & \multirow{2}{*}{Pre-selection (PSC)} & \multirow{2}{*}{Included}\\ \cline{3-4}
 & & $m_N=1~$TeV & $m_N=2.5~$TeV & & \\ 
\hline
$s$-channel & $\mu^+ \mu^- \longrightarrow N_e \Bar{\nu}_e$ & 0.0024 pb & 0.00036 pb & PSC & Yes\\
\hline
VBF & $\mu^+ \mu^- \longrightarrow \Bar{\nu}_\mu \nu_\mu N_e \Bar{\nu}_e$ & $0.0008$ pb & $9.2\times10^{-6}$ pb & PSC & Yes\\
\hline
VBF & $\mu^+ \mu^- \longrightarrow \mu^+ \mu^- N_e \Bar{\nu}_e$ & $3.1\times10^{-5}$ pb & $\sim 10^{-8}$ pb & -- & No\\
\hline 
\end{tabular}%
}
\caption{The signal rate for $N_e$ at 3 TeV muon collider. The pre-selection cut is defined as: $p_T(\ell^-)>20~\text{GeV}$ and $|\eta(\ell^-)|<2.5$ where $\ell^-$ comes from $N_\ell$. The cross section includes the charge conjugate process.}
\label{Table:e_flavor_3TeV_signal}
\end{table}

For the second VBF process $\mu^+\mu^- \longrightarrow \mu^+ \mu^- N_e \Bar{\nu}_e$, a pair of (virtual) photons cannot convert to a pair of neutrinos at tree-level. The cross section of a single photon scatters with an incoming muon is suppressed.  We choose to run the full 2-to-4 VBF processes instead of using gauge boson PDF since there is no divergence. 

\begin{table}[H]
\centering
\resizebox{\textwidth}{!}{%
\begin{tabular}{|c|c|c|c|c|}
\hline
Type & Background process & $\sigma$(w. conj. channel)& Pre-selection (PSC) & Included\\
\hline
$t$-channel & $\mu^+ \mu^- \longrightarrow W^+ e^- \bar{\nu}_e$ & 0.034 pb & PSC & Yes\\
\hline
$t$-channel & $\mu^+ \mu^- \longrightarrow Z e^+ e^-$ & $0.014$ pb & PSC \& missing $e^+$ & No\\
\hline
VBF & $\mu^+ \mu^- \longrightarrow \mu^+ \mu^- W^+ e^- \bar{\nu}_e$ & $0.162$ pb & PSC \& missing $\mu^+\mu^-$ & Yes\\
\hline
VBF & $\mu^+ \mu^- \longrightarrow \mu^- \bar{\nu}_\mu W^+ e^- e^+$ & $0.070$ pb & PSC \& missing $\mu^- e^+$  & Yes\\
\hline
VBF & $\mu^+ \mu^- \longrightarrow \Bar{\nu}_\mu \nu_\mu W^+ e^- \Bar{\nu}_e$ & $0.024$ pb & PSC & Yes\\
\hline 
\end{tabular}%
}
\caption{$N_e$ background at 3 TeV. The final results include the conjugate process that doubles the cross-section reported here.}
\label{Table:e_flavor_3TeV_bkg}
\end{table}

For the background, extra care is needed. The first $t$-channel is $\mu^+\mu^- \longrightarrow W^+ e^- \Bar{\nu}_e$ which is the known channel with $t$-channel singularity. As we discussed earlier, we can ignore this divergence due to the small beam size. The primary contribution is extracted through,
\begin{equation}
    \mu^+ \mu^- \longrightarrow W^+ W^-,\quad W^- \longrightarrow e^- \bar{\nu}_e
\end{equation}
For the second $t$-channel process, the main diagram is 
\begin{equation}
     \mu^+ \mu^- \longrightarrow Z Z,\quad Z \longrightarrow e^- \bar{\nu}_e
\end{equation}
For the first VBF process $\mu^+ \mu^- \longrightarrow \mu^+ \mu^- W^+ e^- \Bar{\nu}_e$, the dominant contribution is from the photon PDF. It turns out this subprocess takes the largest part of the background. We have
\begin{equation}
    \gamma^* \gamma^* \longrightarrow W^+ W^-,\quad W^- \longrightarrow e^- \Bar{\nu}_e
\end{equation}
The second largest VBF process is on the second last row, $\mu^+ \mu^- \longrightarrow \mu^- \bar{\nu}_\mu W^+ e^- e^+$. The primary contribution is from the scattering of the virtual photon and $\mu^+$. The $t$-channel singularity shows up in $\gamma^* \mu^+ \longrightarrow \bar{\nu}_\mu W^+ e^- e^+$ due to the subprocess $\gamma^* \nu_e \longrightarrow W^+ e^-$ where $\nu_e$ is the on-shell decay product from $\mu^+$. Hence we generate the following
\begin{equation}
    \gamma^* \mu^+ \longrightarrow \bar{\nu}_\mu W^+ Z,\quad Z \longrightarrow e^+ e^-
\end{equation}
A similar singularity would appear for the last VBF process. We can divide it into the following two channels:
\begin{equation}
\begin{split}
    &\mu^+ \mu^- \longrightarrow \Bar{\nu}_\mu \nu_\mu W^+ W^-,\quad W^- \longrightarrow e^- \Bar{\nu}_e \\
    &\mu^+ \mu^- \longrightarrow Z W^+ W^-,\quad Z \longrightarrow \Bar{\nu}_\mu \nu_\mu,\quad W^- \longrightarrow e^- \Bar{\nu}_e
\end{split}
\end{equation}

%%%%%%%%%%%%%%%%%%%%%%%%%%%%%%%%%%%%%%%%%%%%%%%%%%%%%%%%%%%
\subsection{$e$-flavored HNL at 10 TeV}

The cross sections for $e$-flavored signal processes at 10 TeV muon collider are listed in \autoref{Table:e_flavor_10TeV_signal}. As can be seen, the cross-section drops with increasing $m_N$. The VBF process dominates at low $m_N$, and drops quickly in the high $m_N$ regime due to the PDF suppression. 
%This is because the contribution from the gauge boson PDF concentrates on the low $\hat{s}$ regime. 

\begin{table}[H]
\centering
\resizebox{\textwidth}{!}{%
\begin{tabular}{|c|c|c|c|c|c|}
\hline
\multirow{2}{*}{Type} & \multirow{2}{*}{Signal process} & \multicolumn{2}{c|}{$\sigma/|U_e|^2$ (w. conj. channel)} & \multirow{2}{*}{Pre-selection (PSC)} & \multirow{2}{*}{Included}\\ \cline{3-4}
 & & $m_N=1~$TeV & $m_N=9~$TeV & & \\ 
\hline
$s$-channel & $\mu^+ \mu^- \longrightarrow N_e \Bar{\nu}_e$ & 0.00024 pb & $1.3\times10^{-5}$ pb & PSC & Yes\\
\hline
VBF & $\mu^+ \mu^- \longrightarrow \Bar{\nu}_\mu \nu_\mu N_e \Bar{\nu}_e$ & $0.0046$ pb & $5.2\times10^{-6}$ pb & PSC & Yes\\
\hline
VBF & $\mu^+ \mu^- \longrightarrow \Bar{\nu}_e \nu_e N_e \Bar{\nu}_e$ & $ 6.0\times10^{-6}$ pb & $\sim10^{-9}$ pb & -- & No\\
\hline
VBF & $\mu^+ \mu^- \longrightarrow \mu^+ \mu^- N_e \Bar{\nu}_e$ & $0.00028$ pb & $6.8\times10^{-8}$ pb &
PSC \& missing $\mu^+ \mu^-$ & 
Yes\\
\hline 
\end{tabular}%
}
\caption{The signal rate for $N_e$ at 10 TeV muon collider. The cross section includes the charge conjugate process.}
\label{Table:e_flavor_10TeV_signal}
\end{table}

For 10 TeV $e$-flavored HNL background, the process of $\mu^+ \mu^- \longrightarrow \Bar{\nu}_\mu \nu_\mu W^+ e^- \Bar{\nu}_e$ is negligible comparing to other two VBF processes. As we will see in the next section, after the analysis cuts, $t$-channel is the dominant background for high-$m_N$ regions, and the other two VBF processes dominate in the middle/low-$m_N$ regions.

\begin{table}[H]
\centering
\resizebox{\textwidth}{!}{%
\begin{tabular}{|c|c|c|c|c|}
\hline
Type & Background process & $\sigma$ (w. conj. channel)& Pre-selection (PSC) & Included\\
\hline
$t$-channel & $\mu^+ \mu^- \longrightarrow W^+ e^- \Bar{\nu}_e$ & 0.030 pb & PSC & Yes\\
\hline
$t$-channel & $\mu^+ \mu^- \longrightarrow Z e^+ e^-$ & $0.00010$ pb & PSC missing $e^+$ & No\\
\hline
VBF & $\mu^+ \mu^- \longrightarrow \mu^+ \mu^- W^+ e^- \Bar{\nu}_e$ & $0.40$ pb & PSC \& missing $\mu^- \mu^+$ & Yes\\
\hline
VBF & $\mu^+ \mu^- \longrightarrow \Bar{\nu}_\mu \nu_\mu W^+ e^- \Bar{\nu}_e$ & $0.063$ pb & PSC & No\\
\hline
VBF & $\mu^+ \mu^- \longrightarrow \mu^- \Bar{\nu}_\mu W^+ e^- e^+$ & $0.22$ pb & PSC missing $\mu^- e^+$ & Yes\\
\hline 
\end{tabular}%
}
\caption{$N_e$ background at 10 TeV. The cross section includes the charge conjugate process.}
\label{Table:e_flavor_10TeV_bkg}
\end{table}

%%%%%%%%%%%%%%%%%%%%%%%%%%%%%%%%%%%%%%%%%%%%%%%%%%%%%%%%%%%
%%%%%%%%%%%%%%%%%%%%%%%%%%%%%%%%%%%%%%%%%%%%%%%%%%%%%%%%%%%
%\newpage
\section{Analysis}
\label{sec:analysis}

After generating the signal and background events passing the pre-selection cuts in the previous section, we proceed with the analysis and report in this section. The heavy neutrino mass $m_N$ can be reconstructed through the invariant mass of the hadronic $W$ and $\ell$ (charged lepton) system. We impose mass-window cut on $m_N$ and generally consider the transverse momentum $p_T$, pseudorapidity $\eta$, and energy $E$ of the charged lepton $\ell$ and hadronic weak boson. Selection cuts are chosen to enable a high signal-background ratio while conservatively not too narrow to allow for unaccounted experimental effects. These considerations enable us to derive realistic projections on muon collider physics potential on HNLs.

%%%%%%%%%%%%%%%%%%%%%%%%%%%%%%%%%%%%%%%%%%%%%%%%%%%%%%%%%%%
\subsection{$\mu$-flavored HNL at 3 TeV}

The cuts we have made step by step are listed in the following:

1. Pre-selection cuts:
Only one charged lepton in the final state, satisfying 
$|\eta(\ell)|<2.5, p_T(\ell)>20~\text{GeV}$ and veto additional charged leptons.

2. Central hadronic $W$ selection: 
$|\eta(W)|<2.5, p_T(W)>20~\text{GeV}$, since the forward hadronic W are not detectable (or detectable but have poor resolution and low efficiency).

3. Mass window: The HNL is reconstructed from the combination of $\ell$ and hadronic  $W$. The invariant mass $m_{W\ell}$ is peaked around the $m_N$ value for the signal. Due to finite mass resolution, the mass windows for all benchmarks are set as $m_N\pm5\%m_N$. 

4. Optimization cuts: Customized missing $p_T$ for each $m_N$ and $E(W)<1450~\text{GeV}$.

\begin{table}[th]
\centering
\resizebox{\textwidth}{!}{%
\begin{tabular}{|c|c|c|c|c|}
\hline
Process & Central $W$ & \multicolumn{1}{c|}{\begin{tabular}[c]{@{}c@{}}Mass window\\ 150/500/1500/2500~GeV\end{tabular}} & Optimization\\
\hline
Background & 59.15\% & 0.63/2.5/2.2/1.3\% & 0.31/0.90/0.95/0.62\%\\
$m_N=150$~GeV & 22.09\% & 22.09\% & 21.80\%\\
$m_N=500$~GeV & 91.20\% & 91.20\% & 77.59\%\\
$m_N=1500$~GeV & 99.87\% & 99.87\% & 89.79\%\\
$m_N=2500$~GeV & 99.96\% & 99.96\% & 92.09\%\\
\hline 
\end{tabular}%
}
\caption{Cutflow table for $N_\mu$ at 3 TeV. All processes after pre-selection are set to 100\%. The cutflow table for individual background channels is shown in the appendix in \autoref{Table:mu_flavor_3TeV_cutflow_detail}.
}
\label{Table:mu_flavor_3TeV_cutflow}
\end{table}

We summarize the selection efficiencies after the central $W$ cut, the mass window imposition, and the optimization cuts in \autoref{Table:mu_flavor_3TeV_cutflow}. Starting from the pre-selection events, we calculate the remaining fraction for each step. For low $m_N$, the central $W$ rejects a sizable fraction of the signal events. This low efficiency is caused by the $t$-channel forward and boosted HNL decays into forward $W$s. For higher $m_N$ values, these cuts keep over 90\% signal events. After the mass window cut, the background events are reduced to O(1\%). The optimization can further reject at least half of the background events.

\begin{figure}[th]
    \centering % Not needed
    \includegraphics[width=0.31\textwidth]{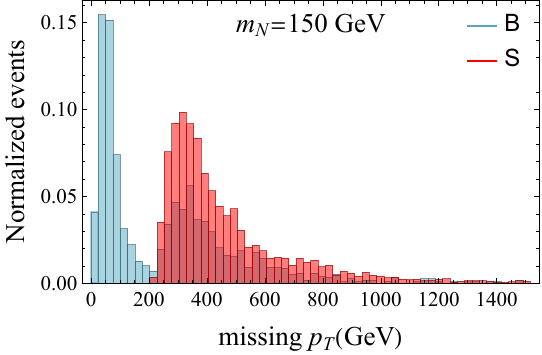}
    \includegraphics[width=0.31\textwidth]{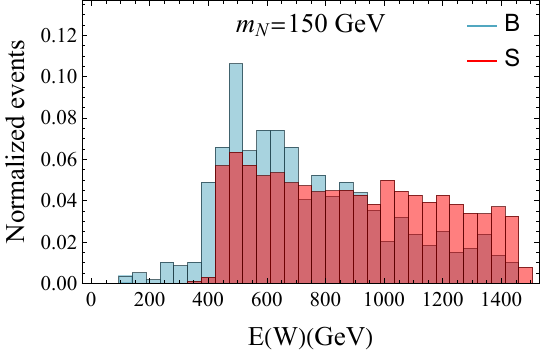}
    \includegraphics[width=0.31\textwidth]{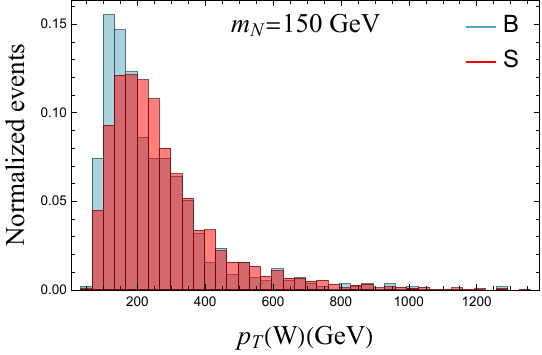}
    \vspace{1ex}
    \includegraphics[width=0.31\textwidth]{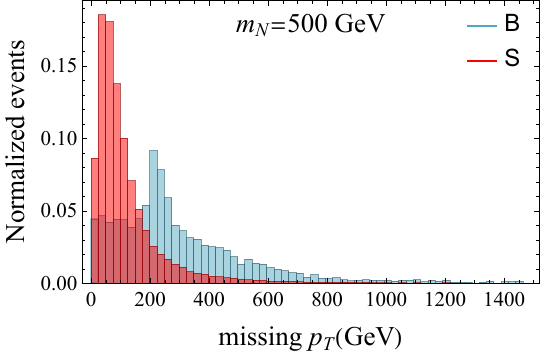}
    \includegraphics[width=0.31\textwidth]{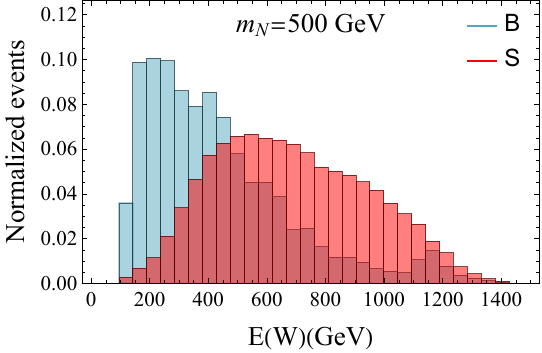}
    \includegraphics[width=0.31\textwidth]{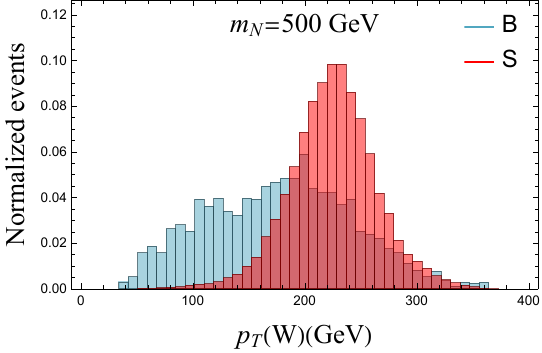}
    \includegraphics[width=0.31\textwidth]{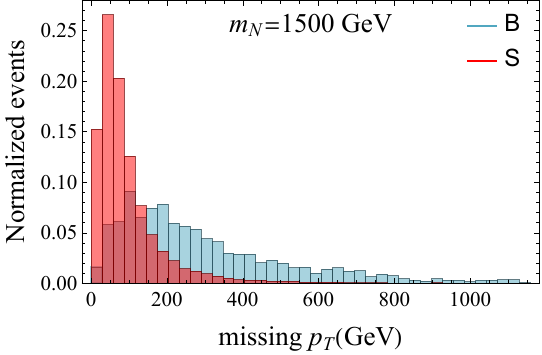}
    \includegraphics[width=0.31\textwidth]{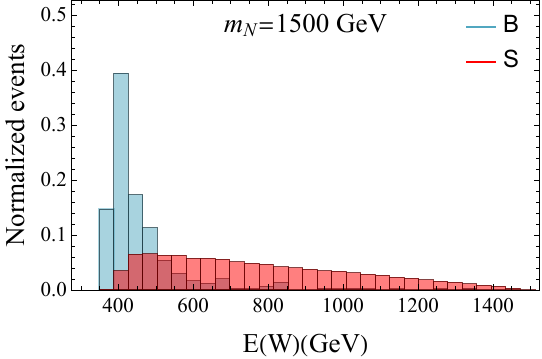}
    \includegraphics[width=0.31\textwidth]{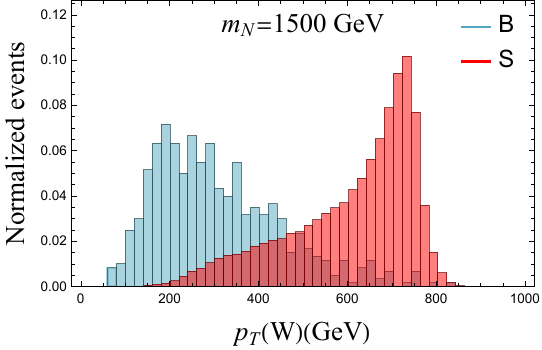}
    \vspace{1ex}
    \includegraphics[width=0.31\textwidth]{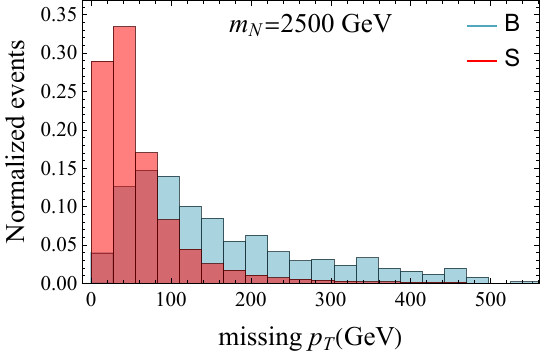}
    \includegraphics[width=0.31\textwidth]{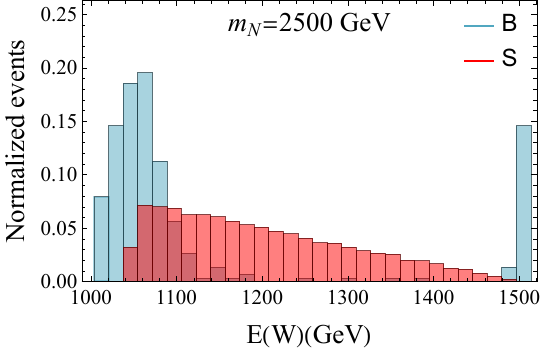}
    \includegraphics[width=0.31\textwidth]{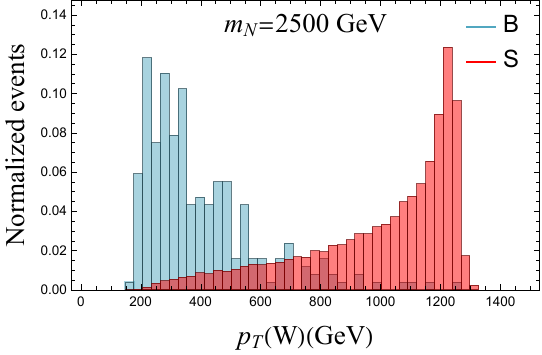}
    \caption{The normalized signal and background distributions $\mu$-flavored HNL at 3 TeV. The left column refers to the missing $p_T$ distribution after pre-selection and central $W$ cut. The middle column refers to $E(W)$ distribution after missing $p_T$ cuts. The right column refers to $p_T(W)$ distribution after $E(W)$ cut. Blue and red bins represent background and signal distributions, respectively.}
    \label{fig:pTW_mu_flavor_3TeV}
\end{figure}

After applying pre-selection, central $W$, and mass-window cuts, the missing $p_T$ cut is implemented. The distribution of missing $p_T$ for four selective $m_N$ benchmark values are shown in the first column of \autoref{fig:pTW_mu_flavor_3TeV}. The red normalized distributions represent the signal events, and the blue represents the background events. For example, we can put a lower cut on $\slashed{p}_T = 200$ GeV for $m_N=150~\text{GeV}$ to reject the majority of background events 
while keeping most of the signal events. For higher $m_N$ values, the signal distribution moves to the lower $\slashed{p}_T$ regime. We can maintain most of the signal events in such cases by imposing an upper cut on $\slashed{p}_T$. For example, we implement the cut $\slashed{p}_T < 200$ GeV for $m_N = 1500$~GeV.

Furthermore, we put a small cut on $E(W)<1450~\text{GeV}$. A majority background events are from $\mu^+ \mu^- \longrightarrow W^+ W^-, W^- \longrightarrow \mu^- \bar{\nu}_\mu$ in which the on-shell $W$ boson carries the energy $E(W^+) = \sqrt{s}/2$. As shown in the middle column of \autoref{fig:pTW_mu_flavor_3TeV}, in the high-mass regions, the background events tend to accumulate at 1500 GeV. For the low-mass region, there is no such pattern, but such a small cut does not affect the background and signal. Generally, both $E(W)$ and $p_T(W)$ provide more distinguishing power in the high-mass region. After imposing one of the above two variables, the distribution of the other variable becomes less distinctive between signal and background. However, the background accumulation at 1500 GeV only appears in $E(W)$ distribution and can be rejected by a light cut on $E(W)$. Therefore, we choose $p_T(W)$ as the final variable to improve the sensitivity after $E(W)<1450$ GeV.

After the optimization cuts, the background and signal can be well separated by the  distribution of W boson's $p_T$, especially in the high-mass regions. As shown on the right column of \autoref{fig:pTW_mu_flavor_3TeV}, the peak value of $p_T(W)$ stays at around 200 GeV, while the peak for signal shifts to higher $p_T$ regime as $m_N$ increases.

Considering the difference between background and signal distribution, we compute the significance by breaking the $p_T(W)$ distribution into 20 bins: $s=\sqrt{\sum S_i^2/B_i}$. The exclusion limit of $|U|^2$ is calculated under 95\% CL ($s<1.96$).

%%%%%%%%%%%%%%%%%%%%%%%%%%%%%%%%%%%%%%%%%%%%%%%%%%%%%%%%%%%%%%
\subsection{$\mu$-flavored HNL at 10 TeV}

The analysis for $\mu$-flavored case at 10 TeV is the same as 3 TeV. After pre-selection, central $W$ and mass-window cuts, the customized cut on missing $p_T$, and a small cut on $E(W)$ are implemented. We replaced the cut $E(W)<1450$ GeV with $E(W)<4950$ GeV. The calculation of significance is changed from 20 bins to 2 bins due to low statistics. The cutflow table is shown in \autoref{Table:mu_flavor_10TeV_cutflow}.

\begin{table}[th]
\centering
\resizebox{\textwidth}{!}{%
\begin{tabular}{|c|c|c|c|c|}
\hline
Process & Central $W$ & \multicolumn{1}{c|}{\begin{tabular}[c]{@{}c@{}}Mass window\\ 150/1000/5000/9000~GeV\end{tabular}} & Optimization\\
\hline
Background & 34.33\% & 0.65/0.79/0.66/0.33\% & 0.057/0.37/0.22/0.16\%\\
$m_N=150$~GeV & 55.04\% & 55.04\% & 55.04\%\\
$m_N=1000$~GeV & 54.75\% & 54.75\% & 51.63\%\\
$m_N=5000$~GeV & 99.93\% & 99.93\% & 97.46\%\\
$m_N=9000$~GeV & 99.99\% & 99.99\% & 98.27\%\\
\hline 
\end{tabular}%
}
\caption{Cutflow table for $N_\mu$ at 10 TeV. All processes after pre-selection are set to 100\%.}
\label{Table:mu_flavor_10TeV_cutflow}
\end{table}

\begin{figure}[th]
    %\centering % Not needed
    \includegraphics[width=0.31\textwidth]{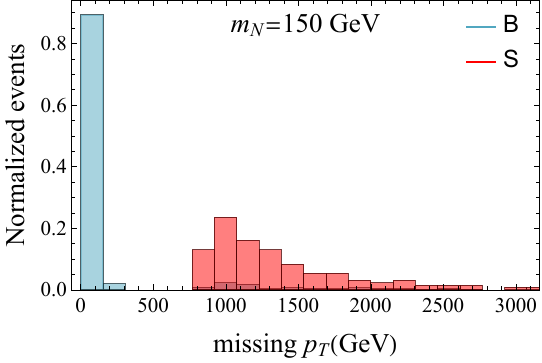}\hfill
    \includegraphics[width=0.31\textwidth]{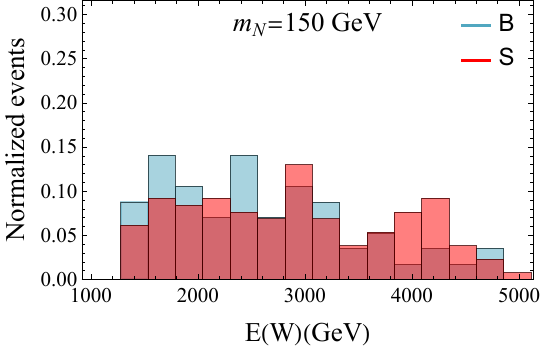}\hfill
    \includegraphics[width=0.31\textwidth]{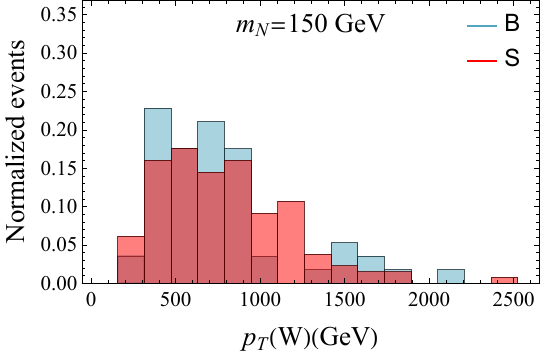}
    \vspace{1ex}
    \includegraphics[width=0.31\textwidth]{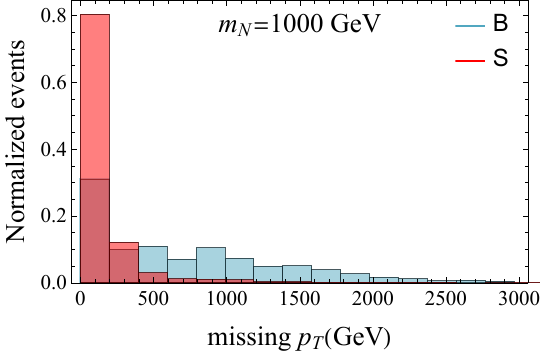}\hfill
    \includegraphics[width=0.31\textwidth]{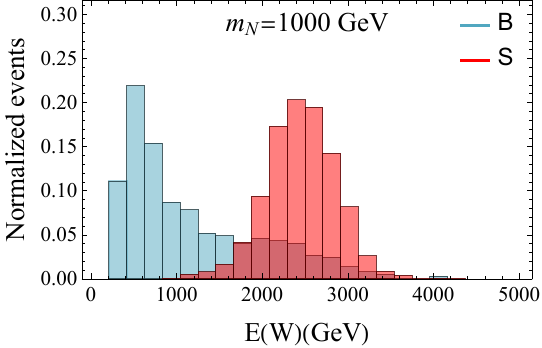}\hfill
    \includegraphics[width=0.31\textwidth]{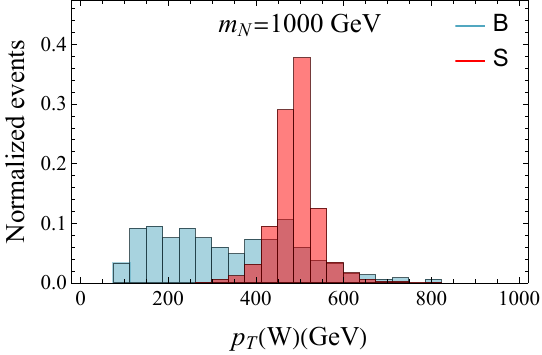}
    \vspace{1ex}
    \includegraphics[width=0.31\textwidth]{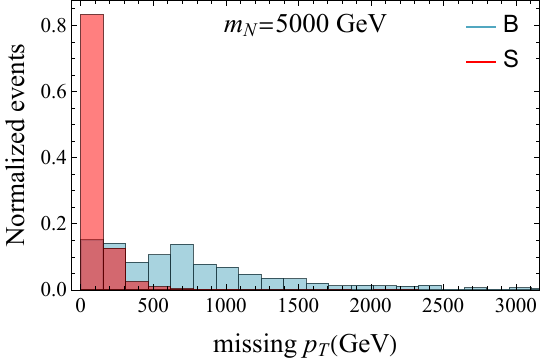}\hfill
    \includegraphics[width=0.31\textwidth]{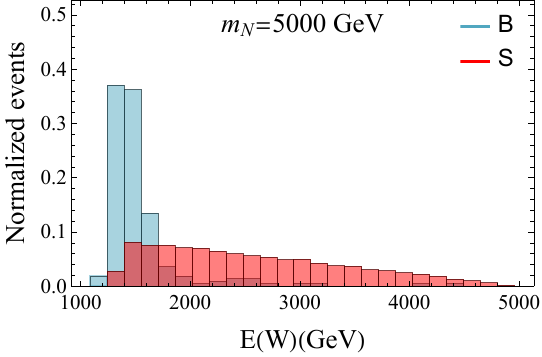}\hfill
    \includegraphics[width=0.32\textwidth]{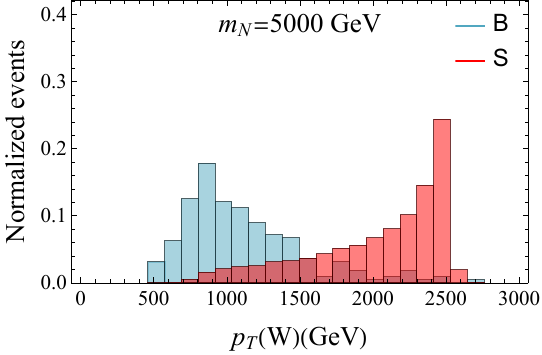}
    \vspace{1ex}
    \includegraphics[width=0.31\textwidth]{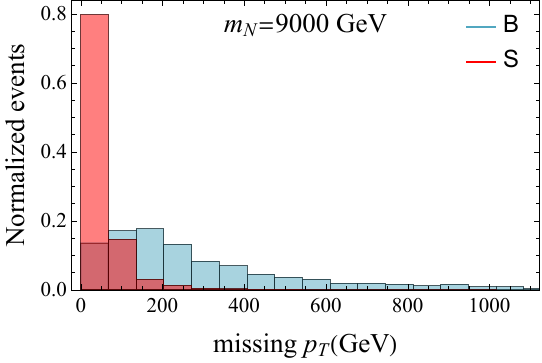}\hfill
    \includegraphics[width=0.31\textwidth]{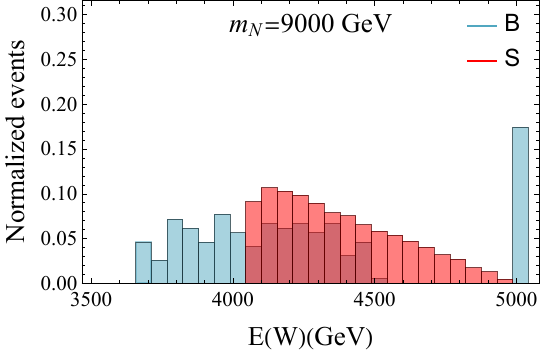}\hfill
    \includegraphics[width=0.32\textwidth]{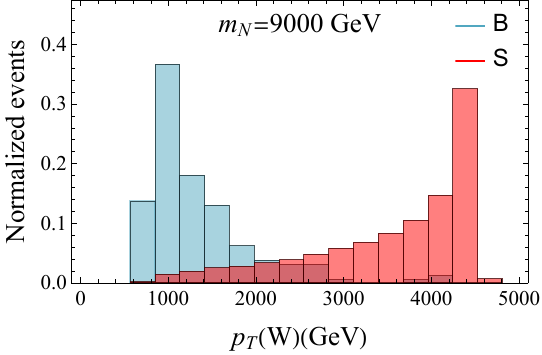}
    \caption{The normalized signal and background distributions of $\mu$-flavored HNL at 10 TeV. The left column refers to the missing $p_T$ distribution after pre-selection and central $W$ cut. The middle column refers to $E(W)$ distribution after missing $p_T$ cuts. The right column refers to $p_T(W)$ distribution after $E(W)$ cut. Blue and red bins represent background and signal distributions, respectively.}
    \label{fig:pTW_mu_flavor_10TeV}
\end{figure}

After pre-selection, central $W$, and mass-window cuts, the optimization cut is implemented on missing transverse momentum $\slashed{p}_T$. The events concentrated at low $\slashed{p}_T$ refer to VBF background. Those spread on higher $\slashed{p}_T$ belong to the non-VBF background as shown in \autoref{fig:pTW_mu_flavor_10TeV}. For $m_N = 150$ GeV, we only retain those events with $\slashed{p}_T < $ 500 GeV. While for higher mass values, we reject the higher $\slashed{p}_T$ events. The cut is conducted on $\slashed{p}_T =$ 500, 400, and 250 GeV for $m_N$ equal to 1000, 1500, and 2500 GeV, respectively.

After applying the cut of missing $p_T$, we add a small cut on $E(W)<4950~\text{GeV}$. As shown in the middle column of \autoref{fig:pTW_mu_flavor_10TeV}, in the high-mass region, the background events tend to accumulate at 5000 GeV. For the low-mass regions, there is no such pattern, but this small cut does not affect the background and signal. For the same reason, in the 3 TeV case, we prefer to leave the cut space for $p_T(W)$, which separates the background better. The $p_T(W)$ distribution is used to compute the significance.

%%%%%%%%%%%%%%%%%%%%%%%%%%%%%%%%%%%%%%%%%%%%%%%%%%%%%%%%%%%%%%
\subsection{$e$-flavored HNL at 3 TeV}

\begin{figure}[th]
    \centering
    %\subfloat[\centering e/tau]
    {{\includegraphics[width=7cm]{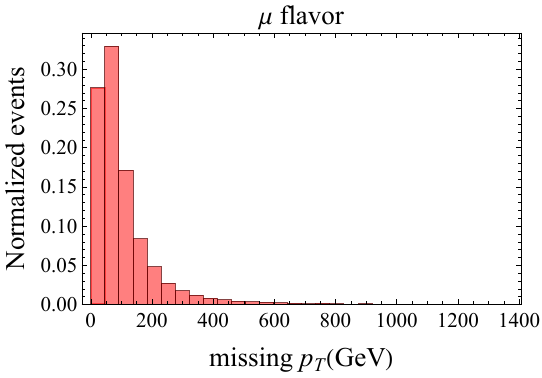} }}%
    \qquad
    %\subfloat[\centering mu]
    {{\includegraphics[width=7cm]{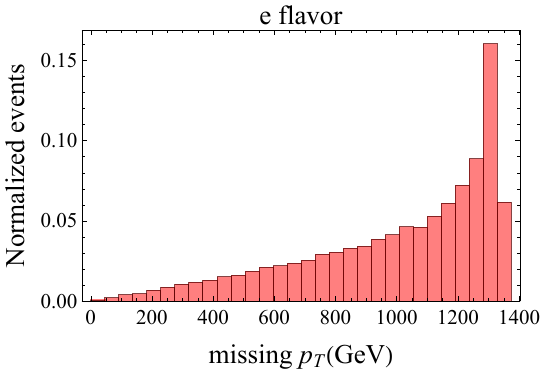} }}%
    \caption{The missing $p_T$ distribution for the leading $\mu^+\mu^- \longrightarrow N_\ell + \bar{\nu}_\ell$ process at $\sqrt{s} = 3$ TeV. The left panel is for $\mu$-flavored signal ($t$-channel), and the right is for $e$-flavored signal ($s$-channel).}
    \label{fig:MPT_s_t_channel}
\end{figure}

The analysis of $e$-flavored HNL at 3 TeV follows $\mu$-flavored analysis. Importantly, the missing $p_T$ distributions differ from the $\mu$-flavored signal. Considering the $\mu$-flavored signal, $t$-channel prefers a forward scattering (low $p_T(\nu)$), and such preference causes a skewed distribution for missing $p_T$, which is the same as the background. However, the $s$-channel is a main part of the $e$-flavored signal, which exhibits an opposite shape. The missing $p_T$ for both channels under $m_N=1000~\text{GeV}$ is shown in \autoref{fig:MPT_s_t_channel}. Therefore, a cut on missing $p_T$ is expected to reduce the background effectively.

The cutflow process followed as $\mu$ case.
We take the same pre-selection cuts, central hadronic $W$ selection, and the mass window cut on $m_{W\ell}$ as in the $\mu$-flavored case. For the optimization cuts, we impose a cut of $E(W) < 1450$ GeV, and the customized $\slashed{p}_T$ cuts for each $m_N$ benchmark.

\begin{table}[th]
\centering
\resizebox{\textwidth}{!}{%
\begin{tabular}{|c|c|c|c|c|}
\hline
Process & Central $W$ & \multicolumn{1}{c|}{\begin{tabular}[c]{@{}c@{}}Mass window\\ 150/500/1500/2500~GeV\end{tabular}} & Optimization\\
\hline
Background & 61.96\% & 1.8/2.2/0.54/2.2\% & 0.066/0.016/0.035/0.051\%\\
$m_N=150$~GeV & 94.52\% & 94.52\% & 70.22\%\\
$m_N=500$~GeV & 97.33\% & 97.33\% & 65.13\%\\
$m_N=1500$~GeV & 98.92\% & 98.92\% & 77.58\%\\
$m_N=2500$~GeV & 98.98\% & 98.98\% & 67.45\%\\
\hline 
\end{tabular}%
}
\caption{Cutflow table for $N_e$ at 3 TeV. All processes after pre-selection are set to 100\%.}
\label{Table:e_flavor_3TeV_cutflow}
\end{table}

We summarize the selection efficiencies under each cut in \autoref{Table:e_flavor_3TeV_cutflow}. After the mass window cuts, only $O(1)\%$ of the background events are maintained, and this ratio can be reduced to $O(0.01)\%$ after the optimization cuts. 

\begin{figure}[th]
    %\centering % Not needed
    \includegraphics[width=0.31\textwidth]{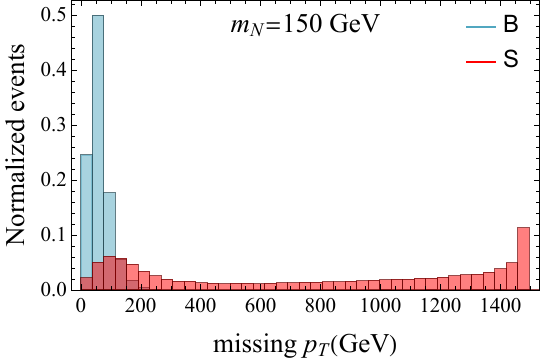}
    \hfill
    \includegraphics[width=0.31\textwidth]{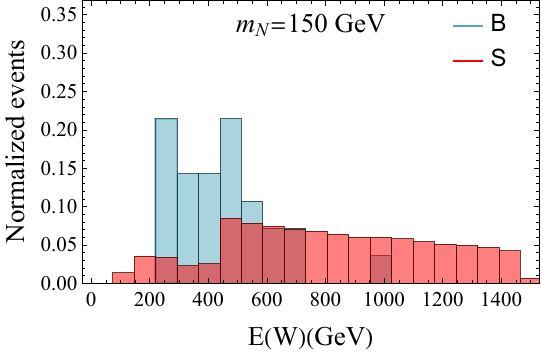}
    \hfill
    \includegraphics[width=0.31\textwidth]{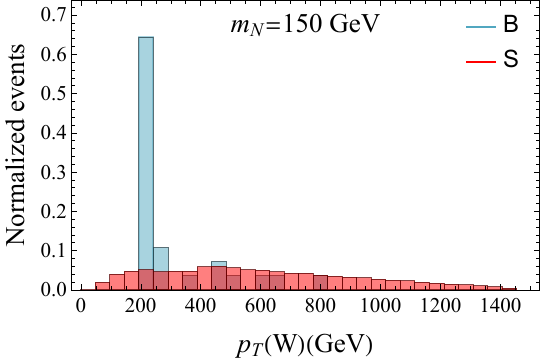}
    \vspace{1ex}
    \includegraphics[width=0.31\textwidth]{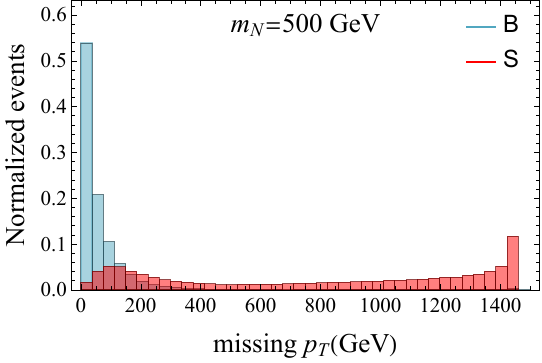}
    \hfill
    \includegraphics[width=0.31\textwidth]{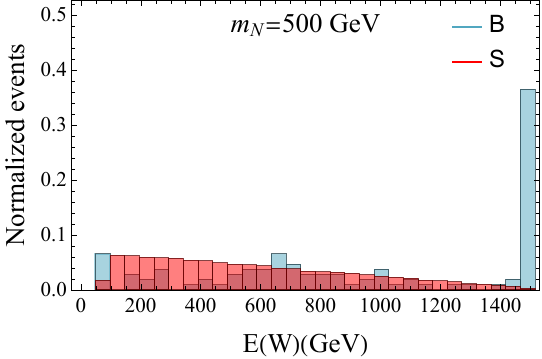}
    \hfill
    \includegraphics[width=0.31\textwidth]{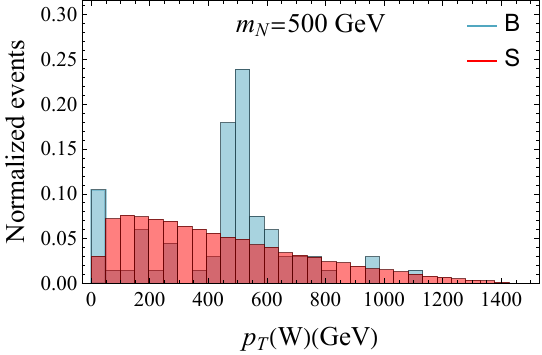}
    \vspace{1ex}
    \includegraphics[width=0.31\textwidth]{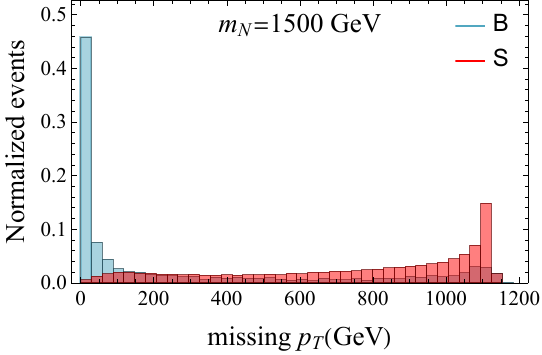}
    \hfill
    \includegraphics[width=0.31\textwidth]{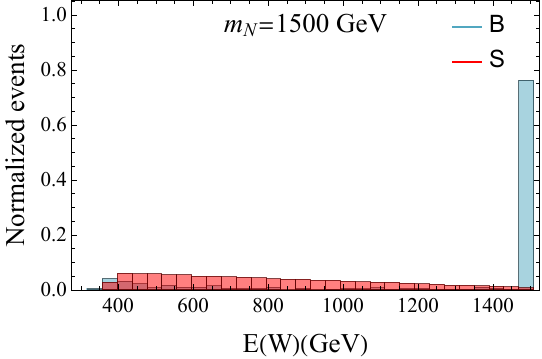}
    \hfill
    \includegraphics[width=0.31\textwidth]{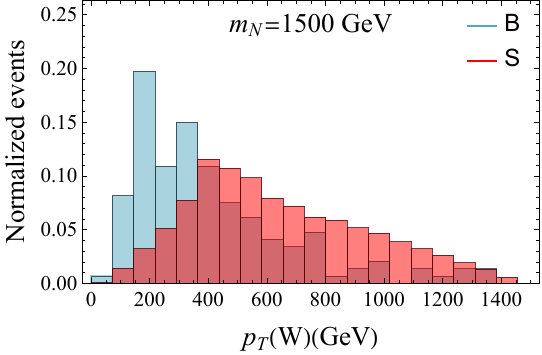}
    \vspace{1ex}
    \includegraphics[width=0.31\textwidth]{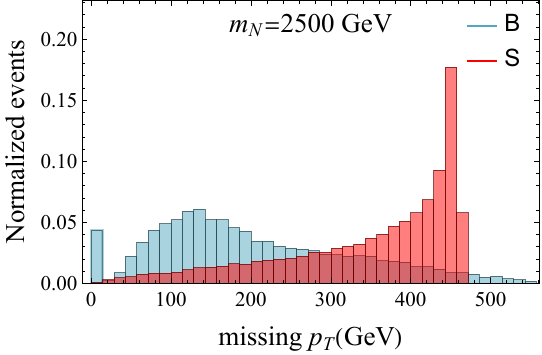}
    \hfill
    \includegraphics[width=0.31\textwidth]{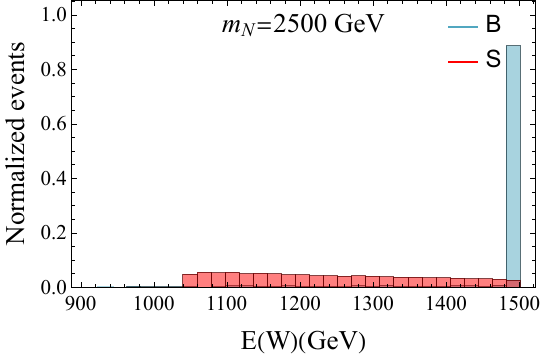}
    \hfill
    \includegraphics[width=0.31\textwidth]{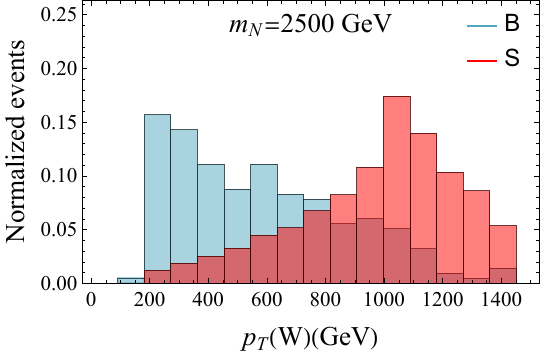}
    %\caption{Comparison4 of large caption dfdf}
    \caption{The normalized signal and background distributions $e$-flavored HNL at 3 TeV. The left column refers to the missing $p_T$ distribution after pre-selection and central $W$ cut. The middle column refers to $E(W)$ distribution after missing $p_T$ cuts. The right column refers to $p_T(W)$ distribution after $E(W)$ cut. Blue and red bins represent background and signal distributions, respectively.}
    \label{fig:pTW_e_flavor_3TeV}
\end{figure}

After the mass window cuts, the $\slashed{p}_T$ distributions are shown on the first column of \autoref{fig:pTW_e_flavor_3TeV}. For low $m_N$ cases, we can see two peaks for the signal shape. The part of signal events that accumulate on the right is from the 2-to-2 $t$-channel process, while the events peaked at low missing $p_T$ regime are from the VBF processes. Since the VBF events are mostly concentrated at low $\sqrt{\hat{s}}$ region, as the value of $m_N$ increases, the VBF contribution to the signal is gradually diminishing. For the background, the dominant channel is $\gamma^* \gamma^* \longrightarrow W^+ W^-,\, W^- \longrightarrow \mu^- \bar{\nu}_e$ which leads to the highly forward $\bar{\nu}_e$. Therefore the background peaks at low $\slashed{p}_T$ region. This feature allows us to select the events at higher $\slashed{p}_T$.

We show the distribution of $E(W)$ and $p_T(W)$ again in the middle and right columns of \autoref{fig:pTW_e_flavor_3TeV}. For a similar reason, we reject the events $E(W)<1450~\text{GeV}$ and finally use the distribution of $p_T(W)$ to compute the sensitivity.

%%%%%%%%%%%%%%%%%%%%%%%%%%%%%%%%%%%%%%%%%%%%%%%%%%%%%%%%%%%%%%
\subsection{$e$-flavored HNL at 10 TeV}

The cutflow process for e-flavored HNL at 10 TeV is similar to 3 TeV in the previous subsection. However, the distribution of missing $p_T$ for the signal is less distinctive from the background than 3 TeV. It is mainly because the primary signal contribution is shifted to the VBF process (\autoref{Table:e_flavor_10TeV_signal}) flattening the distribution.

We take the same pre-selection cuts, central hadronic $W$ selection, and the mass window cut on $m_{W\ell}$ as in the $\mu$-flavored case. For the optimization cuts, we impose the cuts on the hadronic $W$ boson, with $\slashed{p}_T>100~\text{GeV}$ and $E(W)<4950~\text{GeV}$.

\begin{table}[ht]
\centering
\resizebox{\textwidth}{!}{%
\begin{tabular}{|c|c|c|c|}
\hline
Process & Central $W$ & \multicolumn{1}{c|}{\begin{tabular}[c]{@{}c@{}}Mass window\\ 150/1000/5000/9000~GeV\end{tabular}} & Optimization\\
\hline
Background & 34.19\% & 1.2/0.63/0.023/0.134\% & 0.16/0.22/0.011/0.0032\%\\
$m_N=150$~GeV & 83.84\% & 83.84\% & 66.63\%\\
$m_N=1000$~GeV & 93.67\% & 93.67\% & 80.55\%\\
$m_N=5000$~GeV & 99.01\% & 99.01\% & 89.69\%\\
$m_N=9000$~GeV & 99.48\% & 99.48\% & 87.53\%\\
\hline
\end{tabular}%
}
\caption{Cutflow table for $N_e$ at 10 TeV. All processes after pre-selection are set to 100\%.}
\label{Table:e_flavor_10TeV_cutflow}
\end{table}

\begin{figure}[ht]
     \includegraphics[width=0.31\textwidth]{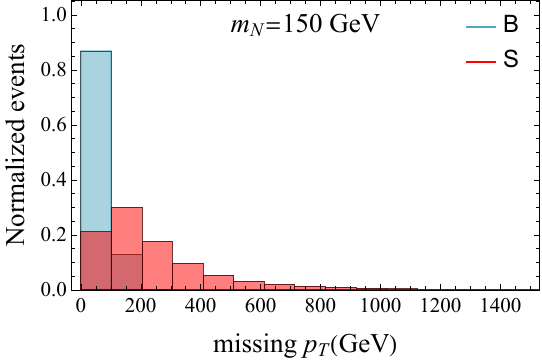}
    \hfill
    \includegraphics[width=0.31\textwidth]{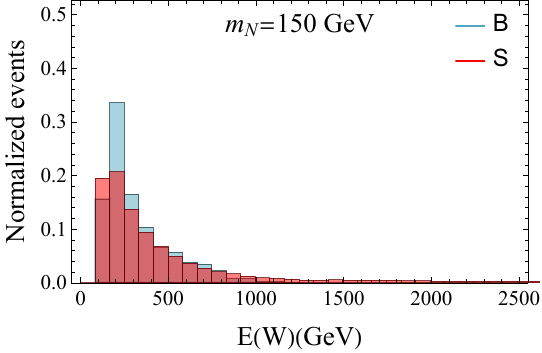}
    \hfill
    \includegraphics[width=0.31\textwidth]{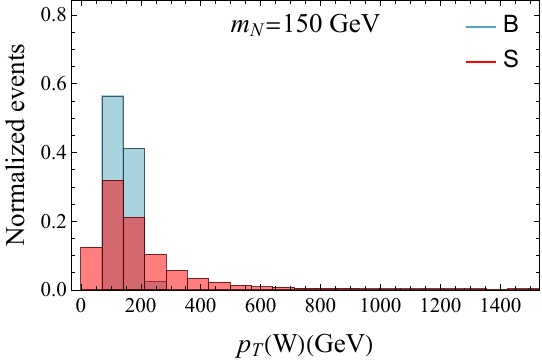}
    \vspace{1ex}
    \includegraphics[width=0.31\textwidth]{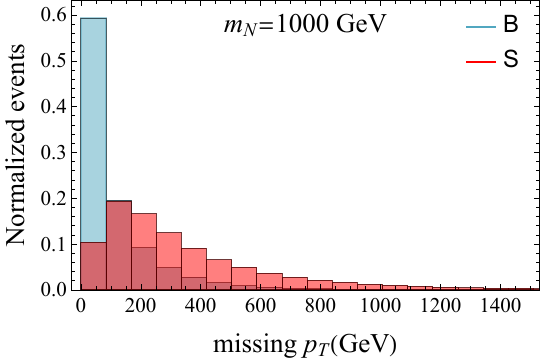}
    \hfill
    \includegraphics[width=0.31\textwidth]{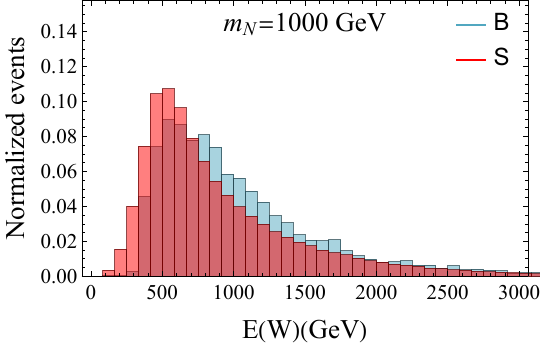}
    \hfill
    \includegraphics[width=0.31\textwidth]{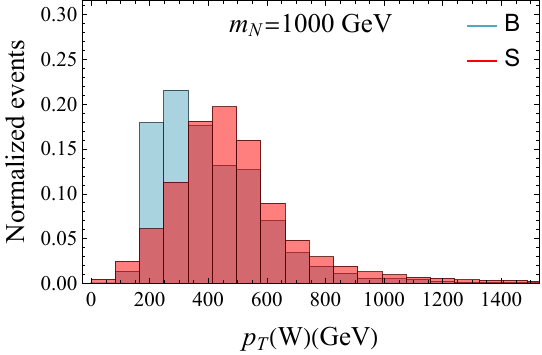}
    \vspace{1ex}
    \includegraphics[width=0.31\textwidth]{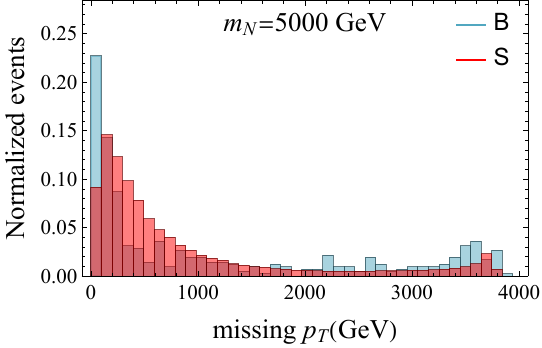}
    \hfill
    \includegraphics[width=0.31\textwidth]{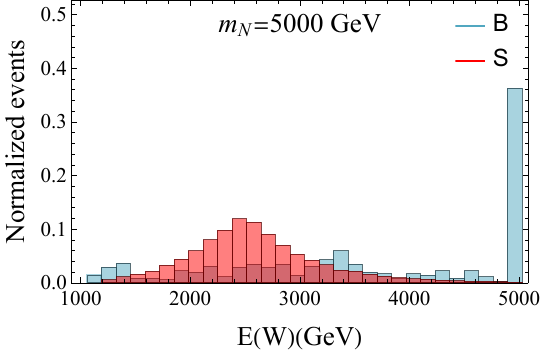}
    \hfill
     \includegraphics[width=0.31\textwidth]{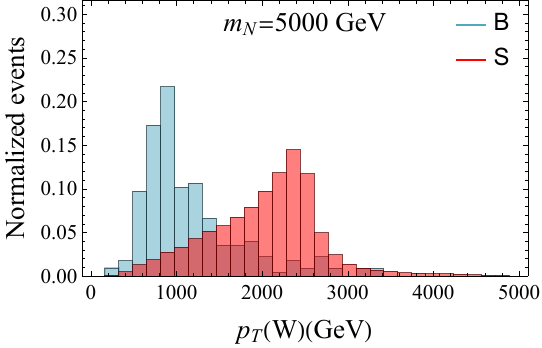}
    \vspace{1ex}
    \includegraphics[width=0.31\textwidth]{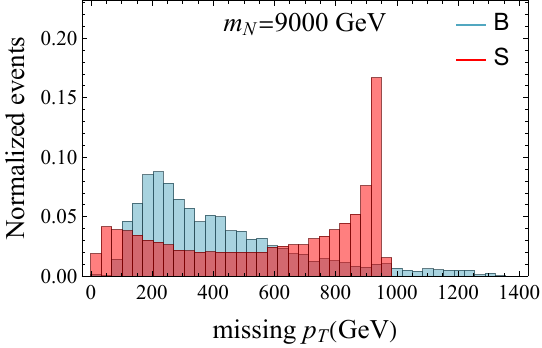}
    \hfill
    \includegraphics[width=0.31\textwidth]{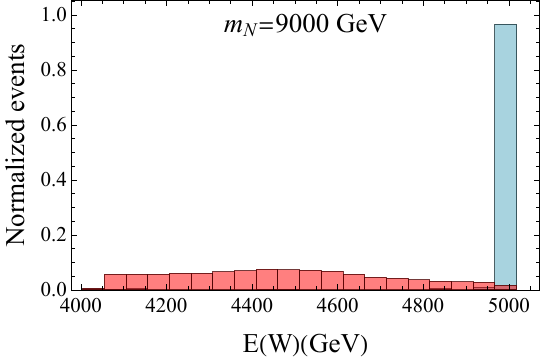}
    \hfill
    \includegraphics[width=0.31\textwidth]{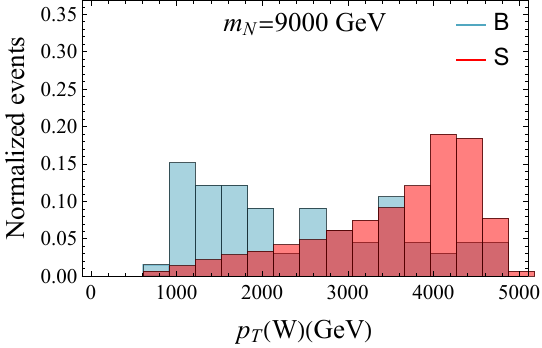}
    %\caption{Comparison4 of large caption dfdf}
    \caption{The normalized signal and background distributions $e$-flavored HNL at 10 TeV. The left column refers to the missing $p_T$ distribution after pre-selection and central $W$ cut. The middle column refers to $E(W)$ distribution after missing $p_T$ cuts. The right column refers to $p_T(W)$ distribution after $E(W)$ cut. Blue and red bins represent background and signal distributions, respectively.}
    \label{fig:pTW_e_flavor_10TeV}
\end{figure}

As shown in the first column of \autoref{fig:pTW_e_flavor_10TeV}, sizable background events tend to be concentrated at low $\slashed{p}_T$ regime. A cut of $\slashed{p}_T>100~\text{GeV}$ is applied except for the high $m_N$ case. 
The following procedures are the same as the previous subsection. 
After a cut on $E(W)$, the $p_T(W)$ distribution 
 as shown in \autoref{fig:pTW_e_flavor_10TeV} is exploited for significance calculation.

\subsection{Projected Sensitivities}
\begin{figure}[htbp]
    \centering
    %\subfloat[\centering e/tau]
    {{\includegraphics[width=0.95\textwidth]{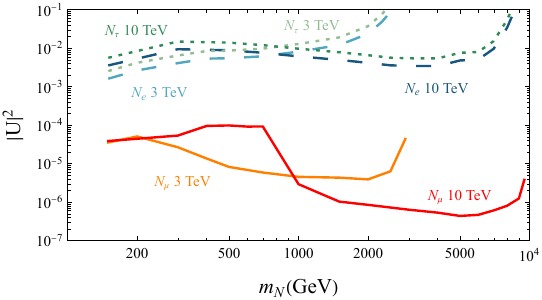} }}%
    \caption{The 95\% projected sensitivity of $|U_\mu|^2$, $|U_e|^2$ and $|U_\tau|^2$ as a function of HNL mass $m_N$ at 3 and 10 TeV muon collider.}
    \label{fig:final-limits-1}%
\end{figure}

After fully implementing the analysis, we evaluate and present the projected sensitivities at muon colliders here. We show the 95\% exclusion limits of $|U_\mu|^2$ and $|U_e|^2$ at muon collider with $\sqrt{s} = 3$ TeV and 10 TeV in \autoref{fig:final-limits-1}. 
 \footnote{Note that for very high $m_N$ value (e.g., $m_N=10$~TeV), with $U^2_\ell > O(0.01)$, the 
coupling of HNL to Higgs boson (parameter $\lambda_\nu$ in \autoref{eq:inverseLag}) approaches the strong coupling regime in the framework of inverse seesaw models. This implies that extra care needs to be provided in the regime $m_N>5$~TeV for the $e$- and $\tau$-flavored HNL, or one studies a model more different from the inverse seesaw model.
%This is not a concern for most of the parameter regimes we are probing and does not our projected sensitivity.
}
The $\mu$-flavored HNL can be well-probed due to $t$-channel signal enhancement. 
%In particular, for $m_N \sim\sqrt{s}/2$, we can achieve the strongest bound. 
The value of $|U_\mu|^2$ can be probed down to $O(10^{-7}) \sim O(10^{-4})$ at 10 TeV and to $O(10^{-6}) \sim O(10^{-4})$ at 3 TeV. For the $e$-flavored case represented by the blue line, the exclusion limit of $|U_e|^2$ is between $O(10^{-3})\%$ to $O(10^{-2})\%$. The $\tau$-flavored HNL has similar signal-background considerations as the electron case. We rescale both the signal and background by 40\%, which accounts for a $\sim$60\% efficiency for the three-prong $\tau$ decays, as a rough estimation of $\tau$-flavor sensitivities in green curves.

Note that for the $\mu$-flavored case, the sensitivities worsen at the low-$m_N$ region. First, this reduction in sensitivity is partially from the lower efficiency in signal detection due to the boosted forward decays discussed earlier (as shown in \autoref{Table:mu_flavor_3TeV_cutflow} and \ref{Table:mu_flavor_10TeV_cutflow}). The other reason is that the missing $p_T$ distribution tends to shift from high missing $p_T$ to low missing $p_T$ as $m_N$ increases, which gradually overlaps with the background (as shown in \autoref{fig:pTW_mu_flavor_10TeV} and \ref{fig:pTW_mu_flavor_10TeV}). On the other hand, as $m_N$ increases, the $p_T(W)$ distribution starts to differ in the background and signal, which improves the sensitivity in high-mass benchmarks.

\begin{figure}[h!]
    \centering
    {{\includegraphics[width=0.95\textwidth]{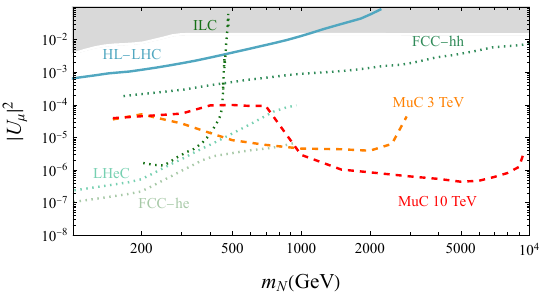} }}%
    \caption{The 95\% exclusion limits on the $|U_\mu|^2$-$m_N$ plane at different experimental facilities including LHC~\cite{Izaguirre:2015pga,Drewes:2019fou,Pascoli:2018heg} and proposed future colliders (LHeC and FCC-he~\cite{Antusch:2019eiz}, FCC-hh~\cite{Pascoli:2018heg,Antusch:2016ejd},ILC~\cite{Mekala:2022cmm,Antusch:2016vyf}). }
    \label{fig:final-limits-all}%
\end{figure}

We also show the projected sensitivities together with LHC and proposed future colliders in \autoref{fig:final-limits-all} for the $\mu$-flavored HNL, which is a benchmark commonly used in future collider studies~\cite{Bose:2022obr,Narain:2022qud,Black:2022cth}. At HL-LHC, we can probe $|U_\mu|^2$ to around $10^{-3}$. The future proton-proton collider could improve $|U_\mu|^2$ by one or two orders of magnitudes. The coverage in HNL mass $m_N$ at LHeC, FCC-he, and ILC are limited by the kinematics from the collision energy. We can see high energy muon colliders uniquely probe large new parameter space on the $|U_\mu|^2-m_N$ plane. 

Note that besides the direct production and search for HNLs, one can also gain knowledge on the HNLs through other precision measurements. For instance, the HNLs mixing with the SM neutrinos changes the $Z$ boson invisible decay width, which dominates the existing constraints on the mixing angle at high HNL mass. For future electroweak factories, such as the FCC-ee~\cite{FCC:2018evy} and CEPC~\cite{CEPCStudyGroup:2018ghi,An:2018dwb,CEPCPhysicsStudyGroup:2022uwl} program, one can also derive complementary constraints, which will be limited by the theoretical uncertainties and systematic uncertainties associated with the precision observables. We anticipate the sensitivity to reach around $10^{-4}$ level from a back-in-the-envelope estimation, and certainly, a full study is in high demand.

%%%%%%%%%%%%%%%%%%%%%%%%%%%%%%%%%%%%%%%%%%%%%%%%%%%%%%%%%%%
%%%%%%%%%%%%%%%%%%%%%%%%%%%%%%%%%%%%%%%%%%%%%%%%%%%%%%%%%%%
%\newpage
\section{Conclusion}
\label{sec:conclusion}

In this paper, we study the unique physics potential of muon colliders in probing heavy neutral leptons. Two benchmark setup for muon colliders are used: $\sqrt{s} = 3$ TeV with integrated luminosity $1~\text{ab}^{-1}$ and  $\sqrt{s} = 10$ TeV with integrated luminosity $10~\text{ab}^{-1}$. For simplicity, we parametrize the HNL in terms of its mass $m_N$ plus the squared mixing angle $|U_\ell|^2$ with the ``active'' SM neutrino. We focus on the regime $m_h < m_N < \sqrt{s}$ where the heavy neutrino promptly decays into massive gauge bosons and Higgs bosons.  For the interesting lower $m_N$ values, we leave it to the future study due to a different consideration for the leading signal production and background composition.

Muon collider is uniquely advantageous for the muon-flavored HNL $N_\mu$ due to the dominance of $t$-channel processes that avoids the $1/s$ suppression. While for $e$- and $\tau$-flavored HNL, the leading 2-to-2 production processes are the $s$-channel diagrams. One also needs to consider the vector boson fusion processes for high-energy muon colliders. To carefully handle the VBF processes, we incorporate the structure functions of the muons, including the EW processes, and compare them with the fixed order calculations in various sub-processes. We adopt different treatments of the VBF processes on a process-by-process basis with corresponding physics considerations. We select the $N_\ell$ decay channel of $N_\ell \rightarrow \ell^- W^+$ followed by hadronic $W$ boson decay. Avoiding the detailed jet analysis, we include the background from both hadronic $W$ and hadronic $Z$ processes and adopt coarse binning and cuts on the hadronic gauge bosons. Therefore our final states are $\ell^- W^+$ plus other particles beyond our detector cut for the signal part and $\ell^- W^+(Z)$ plus ``invisible'' parts.

We adopt a cut-based analysis for HNL signal-background separation on the generated events. After imposing the mass window for the invariant mass of the HNL decay product system, we focus on the distribution of missing transverse momentum, the energy, and the transverse momentum of the hadronically decaying $W$ boson. 
We show the 95\% exclusion limit in the $|U_\ell|^2-m_N$ plane. For the muon-flavored case, future high energy muon colliders can probe $|U_\mu|^2$ down to like $O(10^{-7})-O(10^{-5})$ due to the $t$-channel enhancement. For the electron case, the constraint on $|U_e|^2$ can be probed in a wide mass range for around $10^{-3}$. The results of the tau-flavored case are similar to the results of the electron case.  

While the results in this work already show the promising physics potential for HNLs at future high-energy muon colliders, we have yet to use all the decay channels for the HNLs. For instance, one can study the leptonic decays of the HNLs, new production channels, and also with angular distribution to identify more properties of the HNLs. One can also study the long-lived signatures for HNLs below the tens of the GeV regime. These future directions can further reveal the physics in the neutrino sector and complete our picture of the muon collider potential.

%%%%%%%%%%%%%%%%%%%%%%%%%%%%%%%%%%%%%%%%%%%%%%%%%%%%%%%%%%%
%%%%%%%%%%%%%%%%%%%%%%%%%%%%%%%%%%%%%%%%%%%%%%%%%%%%%%%%%%%
\section*{Acknowledgments}
The authors thank R. Franceschini, S. Pagan Griso, T. Han, L. Li, T. Liu, P. Meade, L.-T. Wang, for their helpful discussions and comments.
This study was supported
in part by the DOE grant DE-SC0022345. Z.L. would like to acknowledge Aspen Center
for Physics for hospitality, supported by National Science Foundation grant PHY1607611. Z.L. also acknowledges the University of Colima, where the final part of this work is completed.
{\flushleft \it Note added:} The preliminary results of this results were reported in the Snowmass2021 reports~\cite{Bose:2022obr,Narain:2022qud} and the muon collider forum report~\cite{Black:2022cth}. With various analysis improvements, the results of our study are presented here. At the final stage of this study, Ref.~\cite{Mekala:2023diu} appeared and showed a consistent result for the $\mu$-flavored HNL at high energy muon collider, using boosted decision tree (BDT). We are also aware of another BDT-based analysis~\cite{Kwok:2023dck} on a related topic, and we acknowledge the authors for coordination.

\appendix

\section{Cutflow table}

In this appendix, we show the detailed cut flow tables for several of our benchmark signals and backgrounds. The cross sections after pre-selections are reported in~\autoref{sec:preselection}, and the optimization analysis follows the descriptions in~\autoref{sec:analysis}. 
\begin{table}[H]
\centering
\resizebox{\textwidth}{!}{%
\begin{tabular}{|c|c|c|c|}
\hline
Background process & Central $W$ & \multicolumn{1}{c|}{\begin{tabular}[c]{@{}c@{}}Mass window\\ 150/1000/5000/9000~GeV\end{tabular}} & Optimization\\
\hline
$\mu^+ \mu^- \longrightarrow W^+ e^- \Bar{\nu}_e$ & 96.82\% & 0.010/0.11/1.7/27.51\% & 0.010/0.080/0.43/0.61\%\\
$\mu^+ \mu^- \longrightarrow \mu^+ \mu^- W^+ e^- \Bar{\nu}_e$ & 41.63\% & 1.4/0.77/0.017/0.00081\% & 0.021/0.28/0.011/0.00023\%\\
$\mu^+ \mu^- \longrightarrow \mu^- \Bar{\nu}_\mu W^+ e^- e^+$ & 17.69\% & 0.39/0.45/0.013/0.00028\% & 0.045/0.15/0.0067/0.00014\%\\
\hline
$\mu^+ \mu^- \longrightarrow N_e \Bar{\nu}_e$ & Central $W$ & Mass window & Optimization\\
$m_N=150$~GeV & 99.62\% & 99.62\% & 99.16\%\\
$m_N=1000$~GeV & 98.05\% & 98.05\% & 97.99\%\\
$m_N=5000$~GeV & 98.94\% & 98.94\% & 98.53\%\\
$m_N=9000$~GeV & 99.35\% & 99.35\% & 95.58\%\\
\hline
$\mu^+ \mu^- \longrightarrow \Bar{\nu}_\mu \nu_\mu N_e \Bar{\nu}_e$ & Central $W$ & Mass window & Optimization\\
$m_N=150$~GeV & 83.34\% & 83.34\% & 64.97\%\\
$m_N=1000$~GeV & 93.48\% & 93.48\% & 79.67\%\\
$m_N=5000$~GeV & 99.02\% & 99.02\% & 88.15\%\\
$m_N=9000$~GeV & 99.78\% & 99.78\% & 68.87\%\\
\hline
$\mu^+ \mu^- \longrightarrow \mu^+ \mu^- N_e \Bar{\nu}_e$ & Central $W$ & Mass window & Optimization\\
$m_N=150$~GeV & 77.69\% & 77.69\% & 61.32\%\\
$m_N=1000$~GeV & 93.06\% & 93.06\% & 80.30\%\\
$m_N=5000$~GeV & 99.08\% & 99.08\% & 85.68\%\\
$m_N=9000$~GeV & 100\% & 100\% & 59.81\%\\
\hline 
\end{tabular}%
}
\caption{Cutflow table for $N_e$ at 10 TeV. All processes after pre-selection is set to 100\%.}
\label{Table:e_flavor_10TeV_cutflow_detail}
\end{table}

\quad

\begin{table}[H]
\centering
\resizebox{\textwidth}{!}{%
\begin{tabular}{|c|c|c|c|c|}
\hline
Background process & Central $W$ & \multicolumn{1}{c|}{\begin{tabular}[c]{@{}c@{}}Mass window\\ 150/500/1500/2500~GeV\end{tabular}} & Optimization\\
\hline
$\mu^+ \mu^- \longrightarrow W^+ \mu^- \Bar{\nu}_\mu$ & 90.37\% & 0.56/3.6/3.9/2.4\% & 0.56/1.1/1.7/1.2\%\\
$\mu^+ \mu^- \longrightarrow Z \mu^+ \mu^-$ & 10.40\% & 0/0.79/0.22/0.090\% & 0/0.084/0/0\%\\
$\mu^+ \mu^- \longrightarrow \mu^+ \mu^- W^+ \mu^- \Bar{\nu}_\mu$ & 62.15\% & 2.7/2.6/0.31/0.021\% & 0/2.4/0.30/0.021\%\\
$\mu^+ \mu^- \longrightarrow \Bar{\nu}_\mu \nu_\mu W^+ \mu^- \Bar{\nu}_\mu$ & 86.21\% & 3.9/3.0/0.35/0.42\% & 1.2/1.3/0.070/0.35\%\\
\hline
$\mu^+ \mu^- \longrightarrow N_\mu \Bar{\nu}_\mu$ & Central $W$ & Mass window & Optimization\\
$m_N=150$~GeV & 22.09\% & 22.09\% & 21.80\%\\
$m_N=500$~GeV & 91.20\% & 91.20\% & 77.59\%\\
$m_N=1500$~GeV & 99.87\% & 99.87\% & 89.79\%\\
$m_N=2500$~GeV & 99.96\% & 99.96\% & 92.09\%\\
\hline 
\end{tabular}%
}
\caption{Cutflow table for $N_\mu$ at 3 TeV. All processes after pre-selection is set to 100\%.}
\label{Table:mu_flavor_3TeV_cutflow_detail}
\end{table}

\quad

\quad

\begin{table}[H]
\centering
\resizebox{\textwidth}{!}{%
\begin{tabular}{|c|c|c|c|c|}
\hline
Background process & Central $W$ & \multicolumn{1}{c|}{\begin{tabular}[c]{@{}c@{}}Mass window\\ 150/1000/5000/9000~GeV\end{tabular}} & Optimization\\
\hline
$\mu^+ \mu^- \longrightarrow W^+ \mu^- \Bar{\nu}_\mu$ & 89.14\% & 0.28/2.4/3.2/1.6\% & 0.28/0.42/1.1/0.80\%\\
$\mu^+ \mu^- \longrightarrow Z \mu^+ \mu^-$ & 1.60\% & 0/0.085/0.039/0.016\% & 0/0.051/0/0\%\\
$\mu^+ \mu^- \longrightarrow \mu^+ \mu^- W^+ \mu^- \Bar{\nu}_\mu$ & 43.39\% & 1.6/0.75/0.011/0\% & 0/0.73/0.0083/0\%\\
\hline
$\mu^+ \mu^- \longrightarrow N_\mu \Bar{\nu}_\mu$ & Central $W$ & Mass window & Optimization\\
$m_N=150$~GeV & 55.04\% & 55.04\% & 55.04\%\\
$m_N=1000$~GeV & 54.75\% & 54.75\% & 51.63\%\\
$m_N=5000$~GeV & 99.93\% & 99.93\% & 97.46\%\\
$m_N=9000$~GeV & 99.99\% & 99.99\% & 98.27\%\\
\hline 
\end{tabular}%
}
\caption{Cutflow table for $N_\mu$ at 10 TeV. All processes after pre-selection is set to 100\%.}
\label{Table:mu_flavor_10TeV_cutflow_detail}
\end{table}

\newpage
\quad

\quad

\begin{table}[h]
\centering
\resizebox{\textwidth}{!}{%
\begin{tabular}{|c|c|c|c|c|}
\hline
Background process & Central $W$ & \multicolumn{1}{c|}{\begin{tabular}[c]{@{}c@{}}Mass window\\ 150/500/1500/2500~GeV\end{tabular}} & Optimization\\
\hline
$\mu^+ \mu^- \longrightarrow W^+ e^- \Bar{\nu}_e$ & 92.97\% & 0.014/0.14/1.5/18\% & 0.014/0.056/0.28/0.43\%\\
$\mu^+ \mu^- \longrightarrow \mu^+ \mu^- W^+ e^- \Bar{\nu}_e$ & 64.22\% & 2.9/2.4/0.23/0.0011\% & 0.0088/0.013/0.0017/0\%\\
$\mu^+ \mu^- \longrightarrow \mu^- \Bar{\nu}_\mu W^+ e^- e^+$ & 33.46\% & 0.91/1.7/0.21/0.031\% & 0/0.0069/0.0030/0\%\\
$\mu^+ \mu^- \longrightarrow \Bar{\nu}_\mu \nu_\mu W^+ e^- \Bar{\nu}_e$ & 84.05\% & 0.034/4.4/2.2/1.1\% & 0/0/0/0\%\\
\hline
$\mu^+ \mu^- \longrightarrow N_e \Bar{\nu}_e$ & Central $W$ & Mass window & Optimization\\
$m_N=150$~GeV & 98.80\% & 98.80\% & 97.08\%\\
$m_N=500$~GeV & 97.68\% & 97.68\% & 93.16\%\\
$m_N=1500$~GeV & 98.90\% & 98.90\% & 67.37\%\\
$m_N=2500$~GeV & 98.97\% & 98.97\% & 42.43\%\\
\hline
$\mu^+ \mu^- \longrightarrow \Bar{\nu}_\mu \nu_\mu N_e \Bar{\nu}_e$ & Central $W$ & Mass window & Optimization\\
$m_N=150$~GeV & 87.85\% & 87.85\% & 28.3\%\\
$m_N=500$~GeV & 96.64\% & 96.64\% & 9.52\%\\
$m_N=1500$~GeV & 98.98\% & 98.98\% & 12.3\%\\
$m_N=2500$~GeV & 99.47\% & 99.47\% & 3.09\%\\
\hline 
\end{tabular}%
}
\caption{Cutflow table for $N_e$ at 3 TeV. All processes after pre-selection is set to 100\%.}
\label{Table:e_flavor_3TeV_cutflow_detail}
\end{table}

\section{Helicity Amplitudes}
We show the helicity amplitudes of the 2-to-2 signal production chanel $\mu^- \mu^+ \rightarrow N_\mu \bar{\nu}_\mu$ in the center-of-mass frame. For simplicity, the muon and neutrino masses are set to zero. The subscripts refer to the helicities of the particles $(\mu^-,~\mu^+,~N,~\bar{\nu})$ respectively. We choose the convention that the incoming $\mu^-$ is along the positive $z$ axis direction and the outcoming $N_\mu$ is emitted along the direction with polar angle $\theta$ and azimuthal angle $\phi = 0$. The angular dependent part is expressed in terms of the Wigner D functions $d_{m,m'}^l$\footnote{$d_{1,1}^1=\frac{1+\cos\theta}{2},~d_{1,-1}^1=\frac{1-\cos\theta}{2},~d_{1,0}^1=-\frac{\sin\theta}{\sqrt{2}}$}.

For $s$-channel diagram, the non-zero helicity amplitudes $\mathcal{M}_{(\mu^-,~\mu^+,~N,~\bar{\nu})}$ are given by:
\begin{align}
    &\mathcal{M}^{(s)}_{-+-+}=\frac{-ig^2U_\mu\cos(2\theta_W)}{2\cos^2\theta_W}\,\frac{s}{s-M_Z^2}\sqrt{1-\frac{m_N^2}{s}}\ d_{1,1}^1\\
    &\mathcal{M}^{(s)}_{+--+}=\frac{-ig^2U_\mu\sin^2\theta_W}{\cos^2\theta_W}\,\frac{s}{s-M_Z^2}\sqrt{1-\frac{m_N^2}{s}}\ d_{1,-1}^1\\
    &\mathcal{M}^{(s)}_{+-++}=\frac{ig^2U_\mu\sin^2\theta_W}{\sqrt{2}\cos^2\theta_W}\,\frac{m_N\sqrt{s}}{s-M_Z^2}\sqrt{1-\frac{m_N^2}{s}}\ d_{1,0}^1\\
    &\mathcal{M}^{(s)}_{-+++}=\frac{-ig^2U_\mu\cos(2\theta_W)}{2\sqrt{2}\cos^2\theta_W}\,\frac{m_N\sqrt{s}}{s-M_Z^2}\sqrt{1-\frac{m_N^2}{s}}\ d_{1,0}^1
\end{align}
where $\theta_W$ is the Weinberg angle. 

For $t$-channel diagram, the non-zero helicity amplitudes are written as:
\begin{align}
    &\mathcal{M}^{(t)}_{-+-+}=\frac{2ig^2 U_\mu \cdot s\sqrt{1-\frac{m_N^2}{s}}\cdot d_{1,1}^1}{(s-m_N^2)(\cos\theta-1)-2M_W^2}\\
    &\mathcal{M}^{(t)}_{-+++}=\frac{-\sqrt{2}ig^2 U_\mu \cdot m_N\sqrt{s}\sqrt{1-\frac{m_N^2}{s}}\cdot d_{1,0}^1}{(s-m_N^2)(\cos\theta-1)-2M_W^2}
\end{align}
In the massless $N_\mu$ case, the amplitude for the final helicity $\pm\pm$ states are vanished.

\bibliographystyle{JHEP}
\bibliography{references}

\end{document}